\documentclass[12pt]{article}
\usepackage{amsfonts,amsmath,amssymb,epsf}
\usepackage{graphics} 
\usepackage{ifpdf}
\ifpdf
\usepackage{epstopdf}
\usepackage{hyperref}
\else
\usepackage[hypertex]{hyperref}
\fi

\advance\textheight by \topskip
\newcounter{thm}

\newcommand\sect[1]{\section{#1}\setcounter{equation}0\setcounter{thm}0} 
                   
\newcommand\be            {\begin{equation}}
\newcommand\bea           {\begin{equation}\begin{array}l\displaystyle}
\newcommand\bearll        {\begin{array}{ll}\displaystyle}
\newcommand\ee            {\end{equation}}
\newcommand\eear          {\end{array}}
\newcommand\enl           {\\[1em]\displaystyle}
\newcommand\etb           {&\!\! \displaystyle}
\newcommand\labl[1]       {\label{#1}\ee}
\newcommand\nxt{\noindent\raisebox{.08em}{\rule{.44em}{.44em}}%
\hspace{.4em}}

\newcommand\arxiv[2]      {\href{http://arXiv.org/abs/#1}{#2}}
\newcommand\doi[2]        {\href{http://dx.doi.org/#1}{#2}}

\newcommand\eps           {\varepsilon}
\newcommand\Fs            {\mathsf{F}}
\newcommand\Ga[1]         {\Gamma\hspace*{-1pt}\big(#1\big)}
\newcommand\Gs            {\mathsf{G}}
\newcommand\Hom           {\mathrm{Hom}}
\newcommand\id            {{\rm id}}
\newcommand\one           {{\bf1}}
\newcommand\Rs            {\mathsf{R}}

\newcommand{\jun}{\gamma}

\newcommand\Cb            {\mathbb{C}}
\newcommand\Hb            {\mathbb{H}}

\newcommand\Rb            {\mathbb{R}}
\newcommand\Zb            {\mathbb{Z}}

\newcommand\Cc            {\mathcal{C}}
\newcommand\Hc            {\mathcal{H}}
\newcommand\Ic            {\mathcal{I}}

\begin{document}

\thispagestyle{empty}
\def\thefootnote{\fnsymbol{footnote}}
\begin{flushright}
\href{http://arXiv.org/abs/1004.1909}{1004.1909 [hep-th]}\\
ZMP-HH/10-10\\
Hamburger Beitr\"age zur Mathematik 371
\end{flushright}
\vskip 5.0em
\begin{center}\LARGE
Non-local conserved charges from defects\\
in perturbed conformal field theory
\end{center}
\vskip 4em
\begin{center}\large
  Ingo Runkel\footnote{Email: {\tt ingo.runkel@uni-hamburg.de}}%
\end{center}
\begin{center}
  Department Mathematik, Universit\"at Hamburg\\
  Bundesstra\ss e 55, 20146 Hamburg, Germany
\end{center}
\vskip 1em
\begin{center}
  April 2010
\end{center}
\vskip 4em
\begin{abstract}
Perturbing a Virasoro minimal model by the (1,3) primary bulk field results in an integrable field theory. In this paper, an infinite set of commuting conserved charges is obtained by considering defects: a one-parameter family of perturbed defect operators is given, and it is shown that these operators mutually commute, that they commute with the Hamiltonian of the perturbed CFT, and that they satisfy a T-system functional relation. The formulation in terms of perturbed defects is a modification of the original prescription for such charges by Bazhanov, Lukyanov and Zamolodchikov.
\end{abstract}

\setcounter{footnote}{0}
\def\thefootnote{\arabic{footnote}}

\newpage

\tableofcontents

\sect{Introduction and summary}\label{sec:intro}

Integrability and conformal symmetry are two important tools to obtain non-perturbative results about two-dimensional quantum field theories. 
In the former case the theory has an infinite number of mutually commuting conserved charges, resulting, in particular, in a factorised scattering theory. In the latter case the infinite dimensional Virasoro algebra constrains the space of states and the correlation functions of the theory.

An integrable field theory need not be conformal, but integrable techniques allow to obtain exact information about the conformal field theories (CFTs) describing the infrared and ultraviolet limits \cite{Zamolodchikov:1991vx}. Conversely, a relevant perturbation of a CFT can give rise to an integrable field theory, and it is a non-trivial problem to decide which perturbations have this property. One way to attack this question is to deduce the existence of higher spin conserved currents in the perturbed theory directly from the decomposition of the space of states of the CFT \cite{Zamolodchikov:1989fp}. Another approach is based on so-called non-local conserved charges \cite{Bazhanov:1994ft,Bazhanov:1996aq}. The latter setting turns out to be more powerful because the non-local charges depend on a spectral parameter, which allows one to formulate functional relations and integral equations for their eigenvalues.

The formalism developed in \cite{Bazhanov:1994ft,Bazhanov:1996aq} has a natural interpretation in terms of CFT in the presence of defect lines: the non-locality of the conserved currents is accounted for by the non-locality of fields localised on defect lines, and the functional relations can be obtained by studying the fusion of perturbed defects.

In \cite{Bazhanov:1994ft} the non-local charges and their functional relations at the conformal point itself are given (for the free boson with background charge and for minimal models), and the corresponding results from the point of view of CFT with defect lines have been obtained in \cite{Runkel:2007wd,Manolopoulos:2009np}. In \cite{Bazhanov:1996aq}, non-local charges for perturbed CFTs are described. The aim of the present paper is to provide the corresponding quantities in terms of defect lines. The defect language allows to prove some properties of the non-local charges which were conjectured in \cite{Bazhanov:1996aq}. 

The defects which will provide the non-local conserved charges are relevant perturbations of certain conformally invariant defects. In fact, these perturbations have already been considered for different reasons: in \cite{Fendley:2009gm} they were studied in the special case of the Ising model to describe tunnelling at a point contact in a quantum Hall system, and in \cite{Kormos:2009sk} they were used to search for new types of conformal defects in unitary minimal models; in \cite{Bajnok:2007jg} such defects were analysed for the Lee-Yang model using integrable scattering methods.

One motivation of the studies in \cite{Bazhanov:1994ft,Bazhanov:1996aq} was to gain a better understanding of the {\em local} conserved charges of the model by considering non-local charges at large values of the spectral parameter. From the defect point of view, this amounts to analysing perturbed defects which flow to the identity defect in the infrared; further investigation of this aspect is left to future work.

\medskip

In more detail, the setting and results of this paper are as follows. We will consider the A-series Virasoro minimal models $M(p,p')$ of central charge $c = 1 - 6 \frac{(p-p')^2}{p p'}$, where $p, p' \ge 2$ are coprime integers. To avoid
short distance singularities later on, we restrict ourselves to parameters $p$, $p'$ such that
\be
  t = \frac{p}{p'} < \frac12 \ .
\labl{eq:t-def}
This excludes in particular the unitary minimal models. Denote by $\Ic$ the set of Kac-labels $(r,s)$ for $M(p,p')$ modulo the equivalence relation $(r,s) \sim (p-r,p'-s)$. For $(r,s) \in \Ic$ let $R_{(r,s)}$ be the irreducible highest weight representation of the Virasoro algebra, whose highest weight vector has $L_0$-eigenvalue
\be
  h_{(r,s)} = \frac{(r-st)^2 - (1-t)^2}{4t}  \ .
\ee
The space of states of $M(p,p')$ is simply
\be
  \Hc = \bigoplus_{i \in \Ic} R_i \otimes_\Cb \overline R_i \ ,
\labl{eq:mm-statespace}
where the overline indicates the action of the anti-holomorphic copy of the Virasoro algebra.
Of particular interest here is the primary bulk field $\Phi \in R_{(1,3)}  \otimes_\Cb \overline R_{(1,3)}$, because this is the relevant field by which we will perturb the CFT. Its left and right conformal weight is $(h_{(1,3)},h_{(1,3)})$ with 
\be
  h_{(1,3)} = 2 t - 1 < 0 \ .
\ee
The field $\Phi$ is normalised such that its operator product expansion (OPE) is given by
\cite{Dotsenko:1984ad,Runkel:1998pm}
\be
  \Phi(z) \Phi(w) ~=~ 
  |z-w|^{-4 h_{(1,3)}} \, C_{\Phi\Phi}^{~\one} \cdot \one
  ~+~
  |z-w|^{-2 h_{(1,3)}} \, C_{\Phi\Phi}^{~\Phi} \cdot \Phi(w) ~+~ \text{other fields} \ ,
\labl{eq:Phi-Phi-OPE}
where
\bea
  C_{\Phi\Phi}^{~\one} = \frac{\sin(3 \pi t)}{\sin( \pi t)} \cdot \eta_\Phi^{\,2} \ ,
  \enl
  C_{\Phi\Phi}^{~\Phi} = 
  - \frac{(1{-}2t)(1{-}3t)}{1{-}4t} \, 
  \frac{\Ga{t}^2 \, \Ga{1{-}2t}^2 \, \Ga{4t}}{\Ga{1{-}t}\,\Ga{2t}^2\,\Ga{3t}\,\Ga{2{-}4t}} 
  \cdot \eta_\Phi \ ,
\eear\labl{eq:bulk-structure-constants}
and $\eta_\Phi$ is a normalisation constant which can be adjusted to the reader's favourite convention.

Let us now turn to defect lines. A defect is a line of inhomogeneity on the surface where the bulk fields can have singularities or discontinuities. A defect is characterised by a `defect condition', that is, a boundary condition for the fields at the defect line. 
This is analogous to the situation where the CFT is defined on the upper half plane and the boundary condition determines the behaviour of bulk fields close to the real line. If the world sheet is a disc with boundary the unit circle, the boundary condition gives rise to a boundary state $\langle\!\langle b |$ in the dual of the space of states. The boundary condition is conformal if the boundary state obeys $\langle\!\langle b | (L_n - \bar L_{-n}) = 0$ for all $n$. 
In the same way, a defect line placed on the unit circle gives rise to an operator $D$ from the space of states to itself. The defect is conformal if the operator obeys \cite{Oshikawa:1996dj}
\be
  [ L_n - \bar L_{-n} , D ] = 0
   \quad \text{ for all $n \in \Zb$} \ .
\ee
A special class of solutions to this condition is provided by totally transmitting defects, which satisfy
\be
  [ L_n , D ] = 0 = [ \bar L_n , D ] 
   \quad \text{ for all $n \in \Zb$} \ .
\labl{eq:top-def-Ln-comm}
This means that the holomorphic and anti-holomorphic components $T$ and $\bar T$ of the stress tensor are continuous across the defect line. As a consequence, the defect line is tensionless and can be deformed on the world sheet without affecting the value of correlators, as long as it is not taken across field insertions or other defect lines. For this reason, defects satisfying condition \eqref{eq:top-def-Ln-comm} are also referred to as topological defects \cite{Bachas:2004sy}. 

Elementary topological defects in $M(p,p')$ (i.e.\ defects that cannot be written as superpositions of other defects) are labelled by Kac-labels $a \in \Ic$, and their defect operator $D_a$ can be expressed in terms of the modular $S$-matrix as \cite{Petkova:2000ip}
\be
  D_a = \sum_{i \in \Ic} \frac{S_{ai}}{S_{1i}} \cdot \id_{R_i \otimes_\Cb \overline R_i} \ ,
\labl{eq:Da-operator}
where $1$ denotes the Kac-label $(1,1)$ and $\id_{R_i \otimes_\Cb \overline R_i}$ is the projector onto the sector $R_i \otimes_\Cb \overline R_i$ of $\Hc$.

If the CFT is defined on the upper half plane or on the unit disc, there are fields which are localised on the boundary of the surface, the boundary fields. The same happens for defects, which can carry defect fields. In the case of topological defects, $T$ and $\bar T$ are continuous across the defect line, and so just as for bulk fields, there are two copies of the Virasoro algebra acting on the space of defect fields. In particular, defect fields on topological defects have a left/right conformal weight $(h,\bar h)$. Because defect fields are confined on the defect line, they do not have to be mutually local and are allowed to have non-integral spin $h-\bar h$. 

In this paper we will be concerned with primary defect fields $\phi$ of weight $(h_{(1,3)},0)$ and $\bar\phi$ of weight $(0,h_{(1,3)})$. These will be the non-local currents whose integrals will provide the non-local conserved charges. 
The space of defect fields has been computed in \cite{Petkova:2000ip,tft1} and one finds that
there exists a unique (up to scalar multiples) defect field of type $\phi$ and $\bar\phi$ on each defect labelled $(r,s)$ with $s = 2,\dots,p'{-}2$. The OPE of these defect fields can be calculated with the methods of \cite{tft4} (see Section~\ref{sec:3dTFT}). Place a defect line labelled $(1,s)$ on the real axis and denote the primary defect field with weights $(h_{(1,3)},0)$ by $\phi^s$ and that with weights  $(0,h_{(1,3)})$ by $\bar\phi^s$. Their OPE reads, for $x>y$,
\bea
  \phi^s(x) \phi^s(y) ~=~ 
  (x{-}y)^{-2 h_{(1,3)}} \,C^{(s)\one}_{\phi\phi} \cdot \one^s ~+~ 
  (x{-}y)^{-h_{(1,3)}} \,C^{(s)\phi}_{\phi\phi} \cdot \phi^s(y) ~+\,\text{other fields}  \ ,
\enl
  \bar\phi^s(x) \bar\phi^s(y) ~=~ 
  (x{-}y)^{-2 h_{(1,3)}}\, C^{(s)\one}_{\phi\phi} \cdot \one^s ~+~ 
  (x{-}y)^{-h_{(1,3)}} \,C^{(s)\phi}_{\phi\phi} \cdot \bar\phi^s(y) ~+\,\text{other fields}  \ ,
\eear\labl{eq:phi-phibar-OPE}
where $\one^s$ is the identity field on the $(1,s)$-defect line. The structure constants
in both OPEs are the same, and they are given by
\bea
   C^{(s)\one}_{\phi\phi} = -  
  \frac{ 1-(s{+}1)t}{1{-}3t}\,
  \frac{
  \Ga{2-(1{+}s)t}\, \Ga{(s{-}1)t} \,\Ga{1{-}t} \,\Ga{2t}\, \Ga{3t}
  }{
  \Ga{(1{+}s)t}\, \Ga{1-(s{-}1)t}\, \Ga{2{-}2t}\, \Ga{t}^2
  }  \cdot \big(\eta_\phi^{(s)}\big)^2 \ ,
   \\[1.5em]\displaystyle
   C^{(s)\phi}_{\phi\phi} =  
    \frac{2 \sin(\pi t) \cos( \pi s t ) }{ \sin(4 \pi t) }\,
  \frac{
  \Ga{2-(s{+}1)t}\, \Ga{(s{-}1)t}}{\Ga{2{-}4t}\, \Ga{2t}} 
   \cdot \eta_\phi^{(s)} \ .
\eear\labl{eq:defect-structure-constants}
Here $\eta_\phi^{(s)}$ is a joint normalisation constant for both $\phi^s$ and $\bar\phi^s$.
Incidentally, the same coefficients also appear in the OPE of the boundary field $\psi^s$ of weight 
$h_{(1,3)}$ on the boundary with Cardy-boundary condition \cite{Cardy:1989ir} labelled $(1,s)$ \cite{Runkel:1998pm},
\be
  \psi^s(x) \psi^s(y) ~=~ 
  (x{-}y)^{-2 h_{(1,3)}} \,C^{(s)\one}_{\phi\phi} \cdot \one^s ~+~ 
  (x{-}y)^{-h_{(1,3)}} \,C^{(s)\phi}_{\phi\phi} \cdot \psi^s(y) ~+\,\text{other fields}  \ .
\ee

Consider the topological defect labelled $(1,s)$ perturbed by the relevant defect field
$\lambda \phi^s + \tilde\lambda \bar\phi^s$, where $\lambda$ and $\tilde\lambda$ are
complex numbers. Denote the operator of the perturbed defect by
\be
  D_s(\lambda,\tilde\lambda) \ .
\labl{eq:Ds-lamtlam}
The results of this paper are the following statements about $D_s(\lambda,\tilde\lambda)$.
Set $q = \exp(\pi i t)$ and take the normalisations $\eta_\phi^{(s)}$ to be related as
\be
  \eta_\phi^{(s)} =   
  \frac{ \Ga{t}\Ga{2{-}3t} }{
         \Ga{(s{-}1)t}\Ga{2{-}(s{+}1)t} } \cdot
 \eta_\phi^{(2)} \ .
\labl{eq:eta-s-relation}
\begin{enumerate}
\item For $m,n = 2,\dots,p'-2$ we have
\be
  \big[ \, D_m(\lambda,\tilde\lambda) \,,\, D_n(\mu,\tilde\mu) \,\big] ~=~ 0
  \quad \text{if} \quad \lambda \cdot \tilde\lambda = \mu \cdot \tilde \mu \ .
\labl{eq:DmDn-comm}
\item 
For $s=2,\dots,p'-2$ and $\eps = \pm 1$,
\bea
  D_2(\lambda,\tilde\lambda) \,
  D_s(q^{s\eps} \lambda, q^{-s\eps} \tilde\lambda) 
\enl
  \quad =~ 
  D_{s-1}(q^{(s+1)\eps} \lambda, q^{-(s+1)\eps} \tilde\lambda) 
  ~+~ 
  D_{s+1}(q^{(s-1)\eps} \lambda, q^{-(s-1)\eps} \tilde\lambda)  \ .
\eear\labl{eq:D2Ds-composition}
On the right hand side, 
it is understood that $D_1(-,-)$ and $D_{p'-1}(-,-)$, on
which there are no defect fields of weight $(h_{(1,3)},0)$ or $(0,h_{(1,3)})$, stand for
the unperturbed defect operators $D_{(1,1)} = \id$ and 
$D_{(1,p'-1)} = \sum_{(r,s) \in \Ic} (-1)^{rp'+sp+1}\, \id_{R_{(r,s)} \otimes_\Cb \bar R_{(r,s)}}$.
From \eqref{eq:D2Ds-composition} one deduces recursively the T-system functional
relation
\be
  D_s(q \lambda, q^{-1} \tilde\lambda) \,
  D_s(q^{-1} \lambda, q \tilde\lambda) 
  ~ =~ \id +
    D_{s-1}(\lambda,\tilde\lambda) 
    D_{s+1}(\lambda,\tilde\lambda) \ .
\labl{eq:tsys-relation}
Furthermore, we have the reflection symmetry
\be
  D_{(1,p'-1)} D_s(\lambda,\tilde\lambda) = 
  D_s(\lambda,\tilde\lambda) D_{(1,p'-1)} = D_{p'-s}((-1)^p\lambda,(-1)^p\tilde\lambda) \ .
\labl{eq:Dinv-Ds}
In particular, $[D_s(\lambda,\tilde\lambda),D_{(1,p'-1)}]=0$ for all
$\lambda,\tilde\lambda$.

\item Write the Hamiltonian of the CFT perturbed by the relevant bulk field $\Phi$ as
\be
   H_P(\gamma) = L_0 + \bar L_0 - \frac{c}{12}
   + \gamma 
   \int_0^{2\pi} \hspace{-.5em}  \Phi(e^{i \theta}) \, d\theta \ .
\ee
Then
\be 
  \big[ \,  H_P(\xi \lambda \tilde\lambda) \,,\, D_s(\lambda,\tilde\lambda) \,\big] ~=~ 0 
\labl{eq:HP-commutator}
for all $\lambda, \tilde\lambda \in \Cb$, and the constant $\xi$ is given by
\be
  \xi = \frac{\big(\eta^{(2)}_\phi\big)^2}{\eta_\Phi} \, \frac{-1}{\sin(3\pi t)} \,
  \frac{\Ga{2{-}3t}\Ga{1{-}t}}{\Ga{2{-}2t}\Ga{1{-}2t}} \ .
\labl{eq:HP-commutator-xi}

\item Let $|(1,1)\rangle\!\rangle$ be the Cardy-boundary state of the vacuum
representation, and let $|(1,s)+\gamma \cdot \psi^s\rangle\!\rangle$ be
the $(1,s)$-boundary state with $\exp(\gamma \int \psi^s)$ 
inserted on the boundary. Then
\be
  D_s(\lambda,\tilde\lambda) \, |(1,1)\rangle\!\rangle = 
  |(1,s)+(\lambda{+}\tilde\lambda) \cdot \psi^s\rangle\!\rangle \ .
\ee
\end{enumerate}
For $\tilde\lambda=\tilde\mu=0$, points 1--4 reduce to the findings of \cite{Runkel:2007wd}. The result of point 3 -- that the non-local charges are conserved -- was already conjectured in \cite{Bazhanov:1996aq}\footnote{Note, however, that the expression for the non-local charges in terms of perturbed defects $D_s(\lambda,\tilde\lambda)$ differs from the prescription in \cite[Eqn.\,(3.22)]{Bazhanov:1996aq}; in particular it uses a different ordering of the fields $\phi$ and $\bar\phi$.}. Point 4 explains the observation on the ratio of perturbed $g$-functions in the conclusions of \cite{Dorey:1999cj}, and points 2 and 4 imply the partition function identity conjectured in \cite[Eqn.\,(5.32)]{Dorey:2000eh}. 

\medskip

Strictly speaking, the results of this paper imply that the identities in 1--4 hold as formal power series in $\lambda$ and $\tilde\lambda$ when inserted into arbitrary correlators on the complex plane. In order to turn them into statements about operators depending on complex parameters, two further points have to be addressed. Firstly, the convergence of the formal sum  in $\lambda$ and $\tilde\lambda$ has to be established. Second, a priori $D_s(\lambda,\tilde\lambda)$ will be a linear map $\Hc \rightarrow \overline\Hc$, where $\Hc$ as given in \eqref{eq:mm-statespace} is the {\em direct sum} of $(L_0{+}\bar L_0)$-eigenspaces, and $\overline\Hc$ is the {\em direct product} of $(L_0{+}\bar L_0)$-eigenspaces. 
In \cite{Runkel:2007wd} this was not a problem, because the linear maps $D_s(\lambda,\tilde\lambda)$ respect the $(L_0{+}\bar L_0)$-grading if either $\lambda$ or $\tilde\lambda$ is zero. If both are nonzero, $D_s(\lambda,\tilde\lambda)$ will in general have contributions in an infinite number of $(L_0{+}\bar L_0)$-eigenspaces even if applied to a homogeneous vector $v$.
To speak of operators one would need to find an appropriate completion of $\Hc$ (which is however smaller than $\overline\Hc$) on which $D_s(\lambda,\tilde\lambda)$ acts. It is expected that these points can be resolved for the models treated here. In the present paper, these issues will not be addressed, but I will nonetheless take the liberty to speak of operators and to treat $\lambda$ and $\tilde\lambda$ as complex numbers.

\bigskip

This paper is organised as follows. In Section~\ref{sec:calc-def}, the necessary background on defects is given, in Section~\ref{sec:pert-def}, the properties listed in points 1--4 are established, and in Section~\ref{sec:3dTFT} the identities between defect correlators that require three-dimensional topological field theory are proved. Appendix~\ref{app:mm-data} lists the chiral data of minimal models used in the main text.

\bigskip
\noindent
{\bf Acknowledgements:} I am grateful to 
Zolt\'an Bajnok,
G\'abor Tak\'acs,
J\"org Teschner 
and 
G\'erard Watts 
for helpful discussions, and I thank 
Zolt\'an Bajnok 
and 
Dimitrios Manolopoulos 
for useful comments on a draft of this paper.

\sect{Calculating with defects}\label{sec:calc-def}

\subsection{Correlators with defect lines}

\subsubsection*{Fields for defects} 

Recall from the introduction that we are considering the A-series Virasoro minimal model $M(p,p')$, that $\Ic$ denotes the set of Kac-labels modulo its $\Zb_2$-identification, and that $R_a$ is the irreducible highest weight representation of the Virasoro algebra labelled by $a \in \Ic$. The following abbreviations for elements of $\Ic$ will be used:
\be
  1 \equiv (1,1) \quad , \quad
  2 \equiv (1,2) \quad , \quad
  3 \equiv (1,3) ~~ .
\ee

Topological defects of $M(p,p')$ are labelled by representations $R_a$ or direct sums thereof \cite{Petkova:2000ip,Petkova:2001ag,tft1,Frohlich:2006ch}. A defect line labelled by the vacuum representation $R_1$ amounts to not inserting a defect at all. We refer to this defect as the {\em identity defect}. An example of a configuration of fields and defect lines is
\be
  \raisebox{-40pt}{
  \begin{picture}(134,92)
   \put(0,0){\scalebox{.75}{\includegraphics{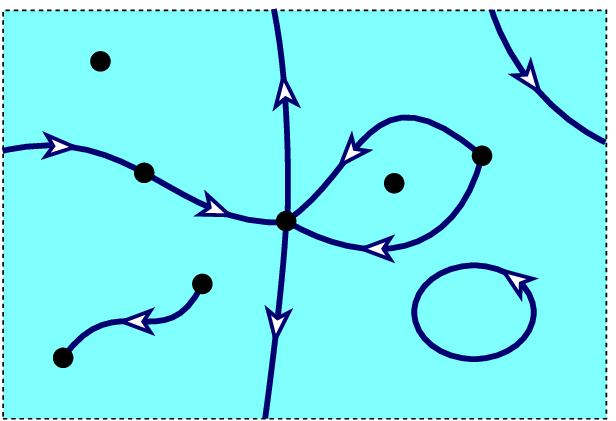}}}
   \put(0,0){
     \setlength{\unitlength}{.75pt}\put(-34,-15){
     \put(46,20) {\scriptsize$ \phi_1 $}
     \put(95,47) {\scriptsize$ \phi_2 $}
     \put(71,73) {\scriptsize$ \phi_3 $}
     \put(65,110) {\scriptsize$ \phi_4 $}
     \put(104,78) {\scriptsize$ \phi_5 $}
     \put(151,77) {\scriptsize$ \phi_6 $}
     \put(176,84) {\scriptsize$ \phi_7 $}
     \put(68,49) {\scriptsize$ a_1 $}
     \put(48,85) {\scriptsize$ a_2 $}
     \put(90,84) {\scriptsize$ a_3 $}
     \put(119,111) {\scriptsize$ a_4 $}
     \put(142,106) {\scriptsize$ a_5 $}
     \put(190,113) {\scriptsize$ a_6 $}
     \put(134,57) {\scriptsize$ a_7 $}
     \put(190,45) {\scriptsize$ a_8 $}
     \put(116,32) {\scriptsize$ a_9 $}
     }\setlength{\unitlength}{1pt}}
  \end{picture}}
  \quad .
\labl{eq:example-configuration}
This picture shows a patch of the complex plane. The general method to express the correlator for such a configuration as a bilinear combination of conformal blocks is given in \cite{tft4,Frohlich:2006ch}, and bits of it will be recalled in Section~\ref{sec:3dTFT}. For now note that defect lines carry an orientation, and that defects can either form closed loops or end at field insertions. 
The field inserted at the junction of multiple defects will be called a {\em junction field}. As special cases one has {\em defect fields}, which are inserted at a junction of two defects with one defect oriented towards the field and one away from it, and {\em disorder fields}, which are located at the start or end of a single defect line. As usual, a {\em bulk field} is a field without any defect lines attached to it. 

Since we are dealing with topological defects, the correlator of the configuration \eqref{eq:example-configuration} does not depend on the precise location of the defect lines. These can be deformed as long as they do not cross each other or field insertions. The fields themselves can in general not be moved without affecting the correlator. An exception are fields of left/right conformal weight $(0,0)$. 

\subsubsection*{Coordinates around field insertions} 

For the correct treatment of defect fields on a defect placed on the unit circle, we need to discuss the local coordinates associated to field insertions. A field insertion $(p,f,\phi)$ consists of a point $p$ on the surface, an injective holomorphic map $f$ (the local coordinate around $p$) from an open neighbourhood of $0 \in \Cb$ to $\Cb$, such that $f(0)=p$, and an element $\phi$ of the corresponding space of (junction, defect, bulk) fields. For correlators on the complex plane one usually implicitly takes $f$ to be $f(\zeta) = \zeta + p$. 
We will only need local coordinates of the form 
\be
  f_{p,\theta}(\zeta) = e^{i \theta} \zeta + p \ ,
\labl{eq:local-coord-theta}
i.e.\ a rotation of the coordinate system by the angle $\theta$ with respect to the standard coordinate on the complex plane. Inside correlators, we have the identity
\be
  (p,f_{p,\theta},\phi) = (p,f_{p,0},e^{i \theta (L_0-\bar L_0)} \phi) \ .
\labl{eq:rotate-local-coord}
For a more general discussion of local coordinates see e.g.\ \cite{Huang1997} or \cite[Sect.\,5.1,\,5.2]{tft4}.

Let us agree on the following convention for local coordinates and then suppress them from the notation in the following. For a bulk field at position $p$ we always choose $f_{p,0}$. For a defect field at position $p$ we consider the normalised tangent vector $\hat t = e^{i\theta}$ to the defect line at $p$ and choose the local coordinate $f_{p,\theta+\pi}$. This means the real axis of the local coordinate is tangent to the defect line but has opposite orientation\footnote{
  This is a convention; it has its origin in orientation choices made in the description
  of  CFT correlators via three-dimensional topological field theory 
  \cite{tft1,tft4,Frohlich:2006ch}.}.
We will see an example in Section~\ref{sec:1bulk1def-ex}.

Let us also agree that when writing an OPE $\phi(x)\psi(y)$ of two defect fields $\phi$ and $\psi$, the defect is placed on the real line, with orientation opposite to the real line (so that $x,y \in \Rb$ and the local coordinates for $\phi$ and $\psi$ are $f_{x,0}$ and $f_{y,0}$, respectively), and that $x>y$. For example, the left hand side of the first OPE in \eqref{eq:phi-phibar-OPE} amounts to the configuration
\be
  \raisebox{-32pt}{
  \begin{picture}(123,78)
   \put(0,0){\scalebox{.75}{\includegraphics{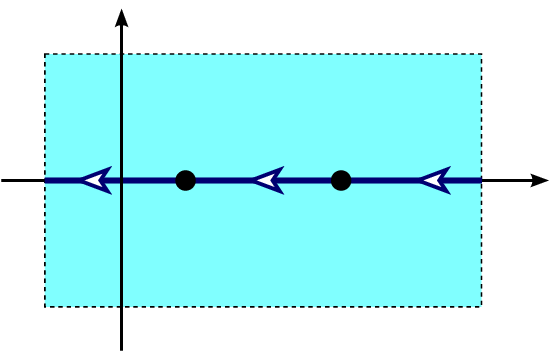}}}
   \put(0,0){
     \setlength{\unitlength}{.75pt}\put(-34,-15){
     \put(142, 72) {\scriptsize$ (1{,}s) $}
     \put( 75, 50) {\scriptsize$ \phi^{s}(y) $}
     \put(120, 50) {\scriptsize$ \phi^{s}(x) $}
     \put(180, 52) {\scriptsize Re}
     \put( 74, 108) {\scriptsize Im}
     }\setlength{\unitlength}{1pt}}
  \end{picture}}
\quad .
\ee
As an aside, the structure constants \eqref{eq:bulk-structure-constants} and \eqref{eq:defect-structure-constants} have simple expressions in terms of $\Fs$-matrices (see \cite{Runkel:1998pm,tft4} and Section~\ref{sec:TFT-struct}),
\be
  C_{\Phi\Phi}^{~\one} = \frac{(\eta_\Phi)^2}{\Fs^{(333)3}_{11}}
  ~~,\quad
  C_{\Phi\Phi}^{~\Phi} = \frac{\eta_\Phi}{\Fs^{(333)3}_{31}}
  ~~,\quad
  C^{(s)\one}_{\phi\phi} = \big(\eta^{(s)}_\phi\big)^2 \, \Fs^{(33s)s}_{s1}
  ~~,\quad
  C^{(s)\phi}_{\phi\phi} = \eta^{(s)}_\phi \, \Fs^{(33s)s}_{s3} \ ,
\labl{eq:structure-const-via-F}
where $s$ stands for $(1,s) \in \Ic$. 
The matrices $\big( \Fs^{(abc)d}_{pq} \big){}_{p,q}$ encode a basis transformation in the space of conformal four-point blocks; some explicit values are given in Appendix~\ref{app:mm-data}.

\subsubsection*{Three-fold junction fields} 

By placing two defect lines labelled by $a$ and $b$ arbitrarily close to each other, one obtains a new topological defect, the {\em fused} topological defect, which we label by $a \star b$. By construction it has preferred three-fold junction fields, namely the identity fields on $a$ and $b$:
\be
  \raisebox{-25pt}{
  \begin{picture}(100,58)
   \put(0,0){\scalebox{.75}{\includegraphics{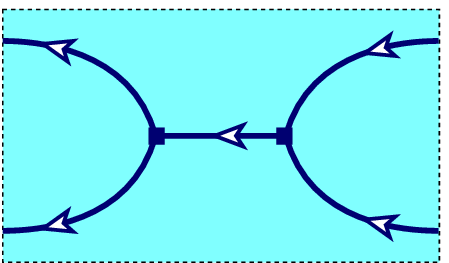}}}
   \put(0,0){
     \setlength{\unitlength}{.75pt}\put(-34,-15){
     \put( 40, 68) {\scriptsize$ b $}
     \put( 40, 30) {\scriptsize$ a $}
     \put(150, 68) {\scriptsize$ b $}
     \put(150, 30) {\scriptsize$ a $}
     \put( 86, 58) {\scriptsize$ a \star b $}
     \put( 66, 48) {\scriptsize$ \one $}
     \put(123, 48) {\scriptsize$ \one $}
     }\setlength{\unitlength}{1pt}}
  \end{picture}}
~=~
  \raisebox{-25pt}{
  \begin{picture}(100,58)
   \put(0,0){\scalebox{.75}{\includegraphics{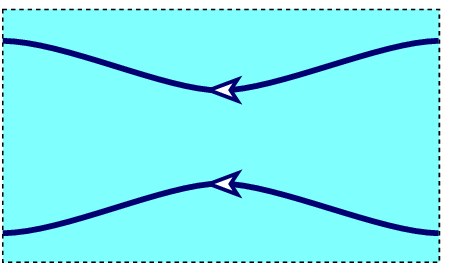}}}
   \put(0,0){
     \setlength{\unitlength}{.75pt}\put(-34,-15){
     \put( 80, 73) {\scriptsize$ b $}
     \put( 80, 26) {\scriptsize$ a $}
     }\setlength{\unitlength}{1pt}}
  \end{picture}}
  \quad .
\ee 
This identity is to be understood as follows. Take two configurations of fields and defects on the complex plane, which are identical outside the patch shown in the pictures, and inside they differ as indicated. Then the correlators of these two configurations will be equal. 

Apart from these canonical junctions, we will need two more types of three-fold junction fields. For $a,b,c \in \Ic$ the space of weight $(0,0)$ junction fields for the configurations
\be
  \raisebox{20pt}{a)}~
  \raisebox{-25pt}{
  \begin{picture}(60,58)
   \put(0,0){\scalebox{.75}{\includegraphics{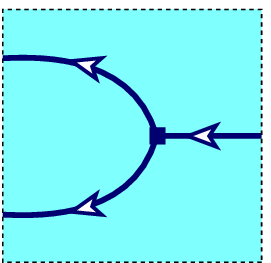}}}
   \put(0,0){
     \setlength{\unitlength}{.75pt}\put(-34,-15){
     \put( 92, 58) {\scriptsize$ c $}
     \put( 41, 77) {\scriptsize$ b $}
     \put (41, 21) {\scriptsize$ a $}
     \put (37, 50) {\scriptsize$ \jun^{a\star b\leftarrow c} $}
     }\setlength{\unitlength}{1pt}}
  \end{picture}}
  \hspace{5em}
  \raisebox{20pt}{b)}~
  \raisebox{-25pt}{
  \begin{picture}(60,58)
   \put(0,0){\scalebox{.75}{\includegraphics{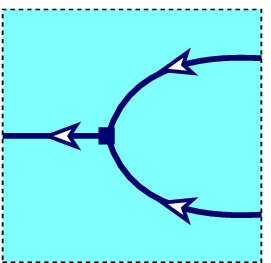}}}
   \put(0,0){
     \setlength{\unitlength}{.75pt}\put(-34,-15){
     \put( 41, 56) {\scriptsize$ c $}
     \put( 92, 77) {\scriptsize$ b $}
     \put (92, 21) {\scriptsize$ a $}
     \put (71, 48) {\scriptsize$ \bar\jun^{c \leftarrow a\star b} $}
     }\setlength{\unitlength}{1pt}}
  \end{picture}}
\labl{eq:3-fold_junction_basis}
has dimension $N_{ab}^{~c}$, i.e.\ the multiplicity of $R_c$ in the fusion product of $R_a$ and $R_b$ \cite{Frohlich:2006ch}. For each choice of $a,b,c \in  \Ic$ with $N_{ab}^{~c}=1$ we pick once and for all a nonzero junction field $\jun^{a\star b\leftarrow c}$ of weight $(0,0)$ for the configuration a) above. For configuration b) we pick a field $\bar\jun^{c \leftarrow a\star b}$ such that
\be
  \raisebox{-25pt}{
  \begin{picture}(100,58)
   \put(0,0){\scalebox{.75}{\includegraphics{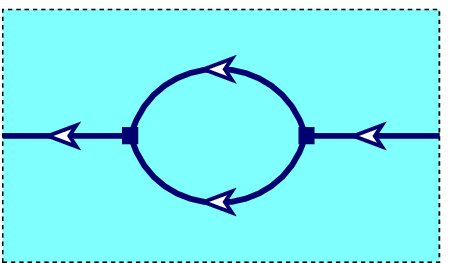}}}
   \put(0,0){
     \setlength{\unitlength}{.75pt}\put(-34,-15){
     \put( 41, 56) {\scriptsize$ d $}
     \put(149, 56) {\scriptsize$ c $}
     \put(112, 71) {\scriptsize$ b $}
     \put(112, 30) {\scriptsize$ a $}
     \put( 78, 48) {\scriptsize$ \bar\jun $}
     \put(110, 48) {\scriptsize$ \jun $}
     }\setlength{\unitlength}{1pt}}
  \end{picture}}
~=~ \delta_{c,d} \, N_{ab}^{~c} ~
  \raisebox{-25pt}{
  \begin{picture}(100,58)
   \put(0,0){\scalebox{.75}{\includegraphics{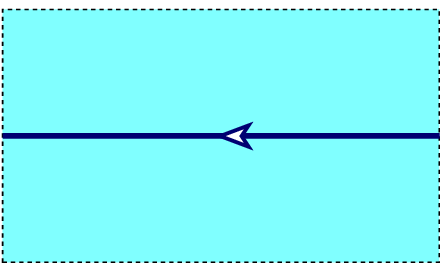}}}
   \put(0,0){
     \setlength{\unitlength}{.75pt}\put(-34,-15){
     \put( 70, 56) {\scriptsize$ c $}
     }\setlength{\unitlength}{1pt}}
  \end{picture}}
\quad .
\labl{eq:lambda-bar-ON}
That such fields $\bar\jun^{c \leftarrow a\star b}$ can be found follows from the TFT approach \cite{Frohlich:2006ch}, though in this simple case it is also easy to argue directly from the non-degeneracy of two-point correlators. The identities below equally follow from the TFT approach. In this section they are merely listed as facts, a sketch of their derivation can be found in Section~\ref{sec:3dTFT}.

The junction fields $\jun^{a\star b\leftarrow c}$ and $\bar\jun^{c \leftarrow a\star b}$ can also be thought of as defect fields which change the $a \star b$-defect to a $c$-defect and back. 
Since the pictorial notation and the notation for OPEs favour different orientations for the arrows when writing out defect fields, let us agree that for all defect fields the two symbols
\be
  \phi^{a \leftarrow b}
  \qquad \text{and} \qquad
  \phi^{b \rightarrow a}
\ee
denote the same field. In particular,
\be
  \jun^{a\star b\leftarrow c} = \jun^{c \rightarrow a\star b}
  \qquad , \qquad
  \bar\jun^{c \leftarrow a\star b}  = 
  \bar\jun^{a\star b \rightarrow c} \ .
\ee
When written as an OPE, the orthonormality condition 
\eqref{eq:lambda-bar-ON} becomes
\be
  \jun^{c \rightarrow a\star b}(x) \,
  \bar\jun^{a\star b \rightarrow d}(y) = \delta_{c,d} \, \one^c(y) \ ,
\labl{eq:fusion-orthonormality}
where $\one^c$ is the identity field on the $c$-defect. As all fields have weight $(0,0)$, there is no coordinate dependence, but the position variable has been included anyway to emphasise that it is an OPE. In the following equations, the position variable will be dropped for weight $(0,0)$ fields. One has the completeness relation
\be
  \sum_{c \in \Ic}  \bar\jun^{a\star b \rightarrow c} \, \jun^{c \rightarrow a\star b} = \one^{a \star b} \ , 
\labl{eq:fusion-completeness}
where it is understood that  $\jun^{c \rightarrow a\star b}$ and $\bar\jun^{a\star b \rightarrow c}$ are zero if the thee-fold junction of $a$, $b$, $c$ does not allow for a weight $(0,0)$ junction field, i.e.\ if $N_{ab}^c = 0$. The decomposition of $a \star b$ into fundamental defects $c$ gives rise to the fusion algebra of topological defects \cite{Petkova:2000ip,Petkova:2001ag,Chui:2001kw,tft1,Frohlich:2006ch}. 

The connectivity of defect networks can be changed via the following identities
\bea
  \raisebox{-25pt}{
  \begin{picture}(100,58)
   \put(0,0){\scalebox{.75}{\includegraphics{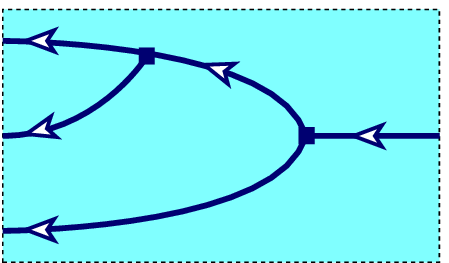}}}
   \put(0,0){
     \setlength{\unitlength}{.75pt}\put(-34,-15){
     \put( 40, 70) {\scriptsize$ c $}
     \put( 60, 48) {\scriptsize$ b $}
     \put( 40, 30) {\scriptsize$ a $}
     \put(149, 56) {\scriptsize$ d $}
     \put(105, 73) {\scriptsize$ p $}
     }\setlength{\unitlength}{1pt}}
  \end{picture}}
~ = ~ \sum_{q \in \Ic} \Fs^{(abc)d}_{pq} ~ 
  \raisebox{-25pt}{
  \begin{picture}(100,58)
   \put(0,0){\scalebox{.75}{\includegraphics{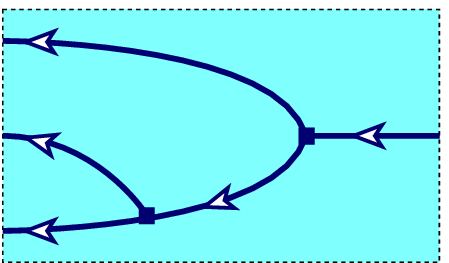}}}
   \put(0,0){
     \setlength{\unitlength}{.75pt}\put(-34,-15){
     \put( 40, 70) {\scriptsize$ c $}
     \put( 60, 48) {\scriptsize$ b $}
     \put( 40, 30) {\scriptsize$ a $}
     \put(149, 56) {\scriptsize$ d $}
     \put(105, 28) {\scriptsize$ q $}
     }\setlength{\unitlength}{1pt}}
  \end{picture}} 
  \quad,
\\[2.5em] \displaystyle
  \raisebox{-25pt}{
  \begin{picture}(100,58)
   \put(0,0){\scalebox{.75}{\includegraphics{pic04b.eps}}}
   \put(0,0){
     \setlength{\unitlength}{.75pt}\put(-34,-15){
     \put( 40, 70) {\scriptsize$ c $}
     \put( 60, 48) {\scriptsize$ b $}
     \put( 40, 30) {\scriptsize$ a $}
     \put(149, 56) {\scriptsize$ d $}
     \put(105, 28) {\scriptsize$ q $}
     }\setlength{\unitlength}{1pt}}
  \end{picture}}
~ = ~ \sum_{p \in \Ic} \Gs^{(abc)d}_{qp} ~ 
  \raisebox{-25pt}{
  \begin{picture}(100,58)
   \put(0,0){\scalebox{.75}{\includegraphics{pic04a.eps}}}
   \put(0,0){
     \setlength{\unitlength}{.75pt}\put(-34,-15){
     \put( 40, 70) {\scriptsize$ c $}
     \put( 60, 48) {\scriptsize$ b $}
     \put( 40, 30) {\scriptsize$ a $}
     \put(149, 56) {\scriptsize$ d $}
     \put(105, 73) {\scriptsize$ p $}
     }\setlength{\unitlength}{1pt}}
  \end{picture}}
  \quad .
\eear\labl{eq:defect-FG-move}
The reason that the $\Fs$-matrices (and their inverses $\Gs$) appear lies in a particular choice made for the $\jun^{x\star y\leftarrow z}$ in the TFT approach, whereby $\jun^{x\star y\leftarrow z}$ is identified with an intertwining operator from $R_x \times R_y$ to $R_z$.

\subsubsection*{Manipulations with defect fields} 

Given $a,b \in \Ic$, a defect field $\phi$ on the $a$-defect and a defect field $\psi$ on the $b$-defect, then $\phi \times \one^b$ and $\one^a \times \psi$ denote defect fields on the fused defect $a \star b$.
The following identities allow to collapse defect bubbles in the presence of the defect fields $\phi^{b\rightarrow a}$ and $\bar\phi^{b\rightarrow a}$ which we take to have left/right conformal weight $(h_f,0)$ and $(0,h_f)$ for some $f \in \Ic$, respectively (see \cite{Runkel:2007wd} and Section~\ref{sec:TFT-struct}):
\be\begin{array}{ll}\displaystyle
  \jun^{d \rightarrow b\star e} ~ (\phi^{b\rightarrow a} \times \one^e)(x) ~ \bar\jun^{a\star e \rightarrow c}
  \etb=~ \frac{\eta^{ab}}{\eta^{cd}} \, \Gs^{(fae)d}_{bc} \cdot \phi^{d \rightarrow c}(x) \ ,
  \enl
  \jun^{d \rightarrow b\star e} ~ (\bar\phi^{b\rightarrow a} \times \one^e)(x) ~ \bar\jun^{a\star e \rightarrow c}
  \etb=~ \frac{\eta^{ab}}{\eta^{cd}} \, \Fs^{(eaf)d}_{bc} \,
  \frac{\Rs^{(eb)d}}{\Rs^{(ea)c}} \cdot \bar\phi^{d \rightarrow c}(x) \ ,
  \enl
  \jun^{d \rightarrow e\star b} ~ (\one^e \times \phi^{b\rightarrow a})(x) ~ \bar\jun^{e\star a \rightarrow c}
  \etb=~ \frac{\eta^{ab}}{\eta^{cd}} \, \Gs^{(fae)d}_{bc} \,
  \frac{\Rs^{(be)d}}{\Rs^{(ae)c}} \cdot \phi^{d \rightarrow c}(x) \ ,
  \enl
  \jun^{d \rightarrow e\star b} ~ (\one^e \times \bar\phi^{b\rightarrow a})(x) ~ \bar\jun^{e\star a \rightarrow c}
  \etb=~ \frac{\eta^{ab}}{\eta^{cd}} \, \Fs^{(eaf)d}_{bc} \cdot \bar\phi^{d \rightarrow c}(x) \ .
\eear\labl{eq:bubble-collapse}
In pictures, the first of these identities reads
\be
  \raisebox{-25pt}{
  \begin{picture}(100,58)
   \put(0,0){\scalebox{.75}{\includegraphics{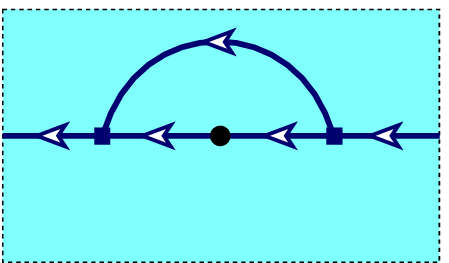}}}
   \put(0,0){
     \setlength{\unitlength}{.75pt}\put(-34,-15){
     \put( 45, 58) {\scriptsize$ c $}
     \put(142, 58) {\scriptsize$ d $}
     \put( 79, 58) {\scriptsize$ a $}
     \put(114, 58) {\scriptsize$ b $}
     \put(117, 76) {\scriptsize$ e $}
     \put( 88, 36) {\scriptsize$ \phi^{a \leftarrow b} $}
     }\setlength{\unitlength}{1pt}}
  \end{picture}}
~=~ \frac{\eta^{ab}}{\eta^{cd}} \, \Gs^{(fae)d}_{bc} ~
  \raisebox{-25pt}{
  \begin{picture}(100,58)
   \put(0,0){\scalebox{.75}{\includegraphics{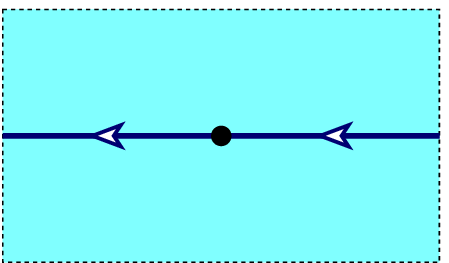}}}
   \put(0,0){
     \setlength{\unitlength}{.75pt}\put(-34,-15){
     \put( 45, 56) {\scriptsize$ c $}
     \put(142, 56) {\scriptsize$ d $}
     \put( 88, 36) {\scriptsize$ \phi^{c \leftarrow d} $}
     }\setlength{\unitlength}{1pt}}
  \end{picture}}   
  \quad .
\ee
The $\eta^{xy}$ are normalisation constants for the fields $\phi^{x\leftarrow y}$ and $\bar\phi^{x\leftarrow y}$. 
For $x=y=(1,s)$ and $f=(1,3)$, the fields  $\phi^{x\leftarrow y}$ and $\bar\phi^{x\leftarrow y}$ are just the fields $\phi^s$ and $\bar\phi^s$ from the introduction, and the normalisation constants are related as $\eta^{(s)}_\phi = \eta^{(1,s)(1,s)}$.

\medskip

The two defect fields $\phi^s$ and $\bar\phi^s$ have a regular OPE, and we can define
\be
  \phi\bar\phi^s(x) = \lim_{\eps\searrow 0} \phi^s(x+\eps)\bar\phi^s(x)
  \quad , \qquad
  \bar\phi\phi^s(x) = \lim_{\eps\searrow 0} \bar\phi^s(x+\eps)\phi^s(x) \ ,
\labl{eq:phiphibar-def}
both of which are defect fields on the $(1,s)$-defect. Unfortunately, because of the order of writing OPEs, in terms of pictures this means, e.g. for $\phi\bar\phi^s$,
\be
\lim_{\eps \searrow 0}
  \raisebox{-25pt}{
  \begin{picture}(100,58)
   \put(0,0){\scalebox{.75}{\includegraphics{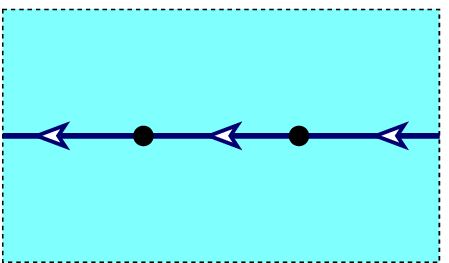}}}
   \put(0,0){
     \setlength{\unitlength}{.75pt}\put(-34,-15){
     \put( 45, 60) {\scriptsize$ (1{,}s) $}
     \put(132, 58) {\scriptsize$ (1{,}s) $}
     \put( 65, 36) {\scriptsize$ \bar\phi^{s}(x) $}
     \put(110, 36) {\scriptsize$ \phi^{s}(x{+}\eps) $}
     }\setlength{\unitlength}{1pt}}
  \end{picture}}
~=~
  \raisebox{-25pt}{
  \begin{picture}(100,58)
   \put(0,0){\scalebox{.75}{\includegraphics{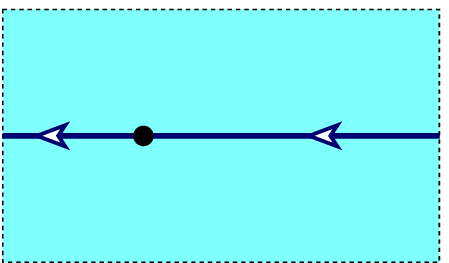}}}
   \put(0,0){
     \setlength{\unitlength}{.75pt}\put(-34,-15){
     \put( 45, 60) {\scriptsize$ (1{,}s) $}
     \put(132, 58) {\scriptsize$ (1{,}s) $}
     \put( 65, 36) {\scriptsize$ \phi\bar\phi^{s}(x) $}
     }\setlength{\unitlength}{1pt}}
  \end{picture}}
  \quad .
\ee
Consider a small defect loop labelled $(1,s)$ with a $\phi\bar\phi^s$ field placed on it. The following identities hold:
\be
  \raisebox{-25pt}{
  \begin{picture}(60,58)
   \put(0,0){\scalebox{.75}{\includegraphics{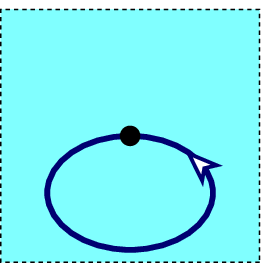}}}
   \put(0,0){
     \setlength{\unitlength}{.75pt}\put(-34,-15){
     \put (58, 25) {\scriptsize$ (1{,}s) $}
     \put (59, 59) {\scriptsize$ \phi\bar\phi^s(x) $}
     }\setlength{\unitlength}{1pt}}
  \end{picture}}
~=~ C_s ~
  \raisebox{-25pt}{
  \begin{picture}(60,58)
   \put(0,0){\scalebox{.75}{\includegraphics{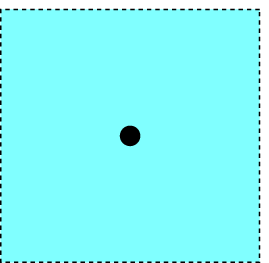}}}
   \put(0,0){
     \setlength{\unitlength}{.75pt}\put(-34,-15){
     \put (77, 50) {\scriptsize$ \Phi(x) $}
     }\setlength{\unitlength}{1pt}}
  \end{picture}}
\qquad , \qquad
  \raisebox{-25pt}{
  \begin{picture}(60,58)
   \put(0,0){\scalebox{.75}{\includegraphics{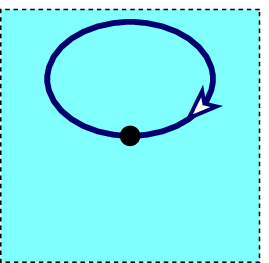}}}
   \put(0,0){
     \setlength{\unitlength}{.75pt}\put(-34,-15){
     \put (58, 73) {\scriptsize$ (1{,}s) $}
     \put (59, 38) {\scriptsize$ \phi\bar\phi^s(x) $}
     }\setlength{\unitlength}{1pt}}
  \end{picture}}
~=~ (C_s)^* \,
  \raisebox{-25pt}{
  \begin{picture}(60,58)
   \put(0,0){\scalebox{.75}{\includegraphics{pic08b.eps}}}
   \put(0,0){
     \setlength{\unitlength}{.75pt}\put(-34,-15){
     \put (77, 50) {\scriptsize$ \Phi(x) $}
     }\setlength{\unitlength}{1pt}}
  \end{picture}}
\quad ,
\labl{eq:phibarphi-to-Phi}
where $(-)^*$ denotes the complex conjugate and
\be
  C_s = \frac{(\eta^{(s)}_\phi)^2}{\eta_\Phi} \cdot e^{\pi i h_3} \, \frac{\dim(s)}{\dim(3)}
  \, \Fs^{(33s)s}_{s1} \ .
\labl{eq:phiphi-circle-coefficient}
Here $s$ stands for $(1,s)$. Even if the tangent to the loop is not horizontal at $x$, there is no factor arising from the change of local coordinates because $L_0-\bar L_0$ acts trivially on $\phi\bar\phi^s$.

\subsection{Example: one bulk and two defect fields}\label{sec:1bulk1def-ex}

Denote by
\be
  D_s[ \psi_1(\theta_1), \dots , \psi_n(\theta_n) ]
\labl{eq:defect-with-fields}
the $(1,s)$-defect placed on the unit circle with clock-wise orientation, with defect field $\psi_k$ inserted at $e^{i \theta_k}$, for $k=1,\dots,n$. Note that by definition the defect field configuration described by \eqref{eq:defect-with-fields} does not depend on the ordering of $\psi_1(\theta_1), \dots, \psi_n(\theta_n)$ in the square brackets. We want to compute the correlator
\be
  f(\alpha,\beta) \cdot \langle \one \rangle = \langle D_s[ \phi^s(\alpha) , \bar\phi^s(\beta) ]\, \Phi(0) \rangle 
  =
  \raisebox{-43pt}{
  \begin{picture}(95,95)
   \put(0,0){\scalebox{.75}{\includegraphics{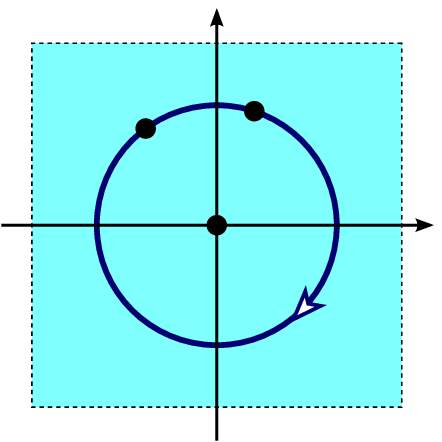}}}
   \put(0,0){
     \setlength{\unitlength}{.75pt}\put(-24,-4){
     \put (90, 56) {\scriptsize$ \Phi(0) $}
     \put (100, 105) {\scriptsize$ \bar\phi^s(e^{i\beta}) $}
     \put (40, 102) {\scriptsize$ \phi^s(e^{i\alpha}) $}
     \put (116, 38) {\scriptsize$ (1{,}s) $}
     }\setlength{\unitlength}{1pt}}
  \end{picture}} 
  \quad ,
\labl{eq:1bulk-1def-corr-def}
which can be expressed as a product of two chiral two-point blocks. To do this we first need to bring the local coordinate of the defect field to standard form. The normalised tangent to the defect at $z = e^{i\alpha}$ is $\hat t = - i e^{i\alpha}$, so that the local coordinate of $\phi^s(z)$ is $f_{z,\alpha+\pi/2}$. We use the identity \eqref{eq:rotate-local-coord} to get
\be
  (z,f_{z,\alpha+\pi/2},\phi^s) = (z,f_{z,0},e^{i (\alpha+\pi/2) h_3 } \phi^s) \ , 
\ee
and similarly, for $w = e^{i\beta}$ and $\bar\phi^s(w)$,
\be
  (w,f_{w,\beta+\pi/2},\bar\phi^s) = (w,f_{w,0},e^{-i (\beta+\pi/2) h_3 } \bar\phi^s) \ .
\ee
In terms of standard coordinates we have the usual expressions for chiral two-point blocks, so that, for some constant $C$,
\be
  f(\alpha,\beta) = C \cdot e^{i (\alpha+\pi/2) h_3 } \,  e^{-i (\beta+\pi/2) h_3 } \, z^{-2 h_3} (w^*)^{-2 h_3} = C \cdot e^{i h_3(\beta-\alpha)} \ .
\ee  
To find the constant $C$ we use the identity \eqref{eq:phibarphi-to-Phi} and the OPE \eqref{eq:Phi-Phi-OPE}:
\bea
  C \langle \one \rangle = 
  \lim_{\eps\searrow 0}
  f(\beta+\eps,\beta) \langle \one \rangle
  =  \langle D_s[ \phi\bar\phi^s(\beta) ] \Phi(0) \rangle  
\enl
=\,
  \raisebox{-43pt}{
  \begin{picture}(95,95)
   \put(0,0){\scalebox{.75}{\includegraphics{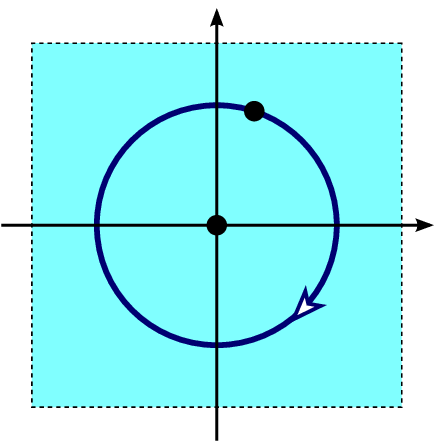}}}
   \put(0,0){
     \setlength{\unitlength}{.75pt}\put(-24,-4){
     \put (90, 56) {\scriptsize$ \Phi(0) $}
     \put (96, 107) {\scriptsize$ \phi\bar\phi^s(e^{i\beta}) $}
     \put (116, 38) {\scriptsize$ (1{,}s) $}
     }\setlength{\unitlength}{1pt}}
  \end{picture}} 
~=\,
  \raisebox{-43pt}{
  \begin{picture}(95,95)
   \put(0,0){\scalebox{.75}{\includegraphics{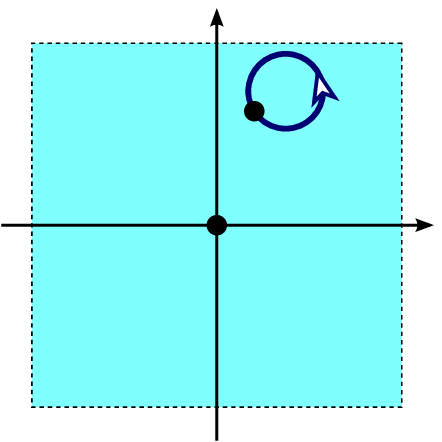}}}
   \put(0,0){
     \setlength{\unitlength}{.75pt}\put(-24,-4){
     \put (90, 56) {\scriptsize$ \Phi(0) $}
     \put (90, 83) {\scriptsize$ \phi\bar\phi^s(e^{i\beta}) $}
     \put (118, 109) {\scriptsize$ (1{,}s) $}
     }\setlength{\unitlength}{1pt}}
  \end{picture}} 
  \,
  = C_s \langle \Phi(e^{i\beta}) \Phi(0) \rangle
  = C_s C_{\Phi\Phi}^{~\one} \langle \one \rangle \ .
\eear\ee
In the first picture we think of the correlator as a correlator on the Riemann sphere with the vacuum inserted at infinity. This allows us to deform the defect loop `through infinity' to arrive at the second picture. Altogether
\be
  f(\beta+\delta,\beta) =  C_s C_{\Phi\Phi}^{~\one} \cdot e^{- i h_3 \delta} =
  \big(\eta^{(s)}_\phi\big)^2 \, \eta_\Phi \cdot 
  e^{i \pi h_3} \dim(s) \, \Fs^{(33s)s}_{s1}\cdot e^{- i h_3 \delta} \ , 
\ee
where $\delta \in {]}0,2\pi{[}$ and $s = (1,s)$. The explicit expressions for the constants
can be found in Appendix~\ref{app:mm-data}.

\subsection{Defect crossings}

As in the introduction, define
\be
  q = \exp( \pi i t) \ ,
\labl{eq:q-def}
with $t = p/p'$. 
Consider the defect labelled by $2 \equiv (1,2) \in \Ic$. The weight $(0,0)$ field $b(\lambda,\mu)$ lives on the fused defect $2 \star 2$ and is defined as
\be
  b_{\lambda,\mu} = (\mu q^{-1} - \lambda q) \cdot \bar\jun^{2\star 2 \rightarrow 1} \jun^{1 \rightarrow 2\star 2} 
  + (\lambda q^{-1} - \mu q) \cdot \bar\jun^{2\star 2 \rightarrow 3} \jun^{3 \rightarrow 2\star 2} \ ,
\labl{eq:b-def}
for $\lambda,\mu \in \Cb$.
As an aside, in the TFT approach, this corresponds to an intertwiner of the representation $R_2 \otimes R_2$ to itself, and one can check that it is given in terms of the braiding isomorphism $c_{R,S} : R \otimes S \rightarrow S \otimes R$ as $\lambda e^{-\pi i t/2} c_{R_2,R_2} - \mu e^{\pi i t/2} c_{R_2,R_2}^{-1}$.

The fields $\phi^2$ and $\bar\phi^2$ on the $2$-defect give rise to the four fields 
$\phi^2{\times}\one^2$, $\bar\phi^2{\times}\one^2$, $\one^2{\times}\phi^2$, $\one^2{\times}\bar\phi^2$ on the $2 \star 2$ defect. 
The following identities obeyed by $b_{\lambda,\mu}$ will be important later when establishing commutativity properties of perturbed defect operators.
\bea
  b_{\lambda,\mu} \,
  \big(\lambda (\phi^2 {\times}\one^2)(x) + \mu (\one^2{\times}\phi^2)(x)\big) 
  = 
  \big(\mu (\phi^2{\times}\one^2)(x) + \lambda (\one^2{\times}\phi^2)(x)\big) \, 
  b_{\lambda,\mu} 
  \ ,
\enl
  b_{\mu,\lambda} \,
  \big(\lambda (\bar\phi^2 {\times}\one^2)(x) + \mu (\one^2{\times}\bar\phi^2)(x)\big) 
  = 
  \big(\mu (\bar\phi^2{\times}\one^2)(x) + \lambda (\one^2{\times}\bar\phi^2)(x)\big) \, 
  b_{\mu,\lambda}
  \ ,
\eear\labl{eq:b-exchange-relations}
where $\lambda,\mu \in \Cb$. Note that the prefactors $\lambda$ and $\mu$ of the two fields get exchanged when taking $b_{\lambda,\mu}$ past the linear combination. The field $b_{\lambda,\mu}$ plays a similar role to that of the $R$-matrix in \cite{Bazhanov:1996aq}. Pictorially, the first identity amounts to
\bea
\lambda \,
  \raisebox{-25pt}{
  \begin{picture}(100,58)
   \put(0,0){\scalebox{.75}{\includegraphics{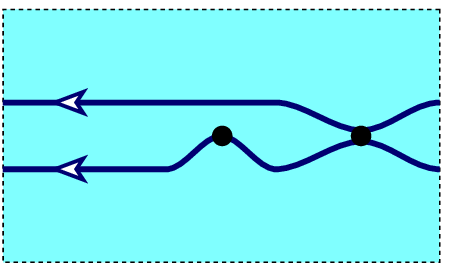}}}
   \put(0,0){
     \setlength{\unitlength}{.75pt}\put(-34,-15){
     \put(60, 67) {\scriptsize$ (1{,}2) $}
     \put(60, 31) {\scriptsize$ (1{,}2) $}
     \put(128, 62) {\scriptsize$ b_{\lambda,\mu} $}
     \put( 92, 37) {\scriptsize$ \phi^2 $}
     }\setlength{\unitlength}{1pt}}
  \end{picture}}
~+~
\mu \,
  \raisebox{-25pt}{
  \begin{picture}(100,58)
   \put(0,0){\scalebox{.75}{\includegraphics{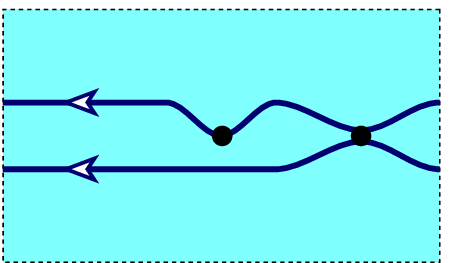}}}
   \put(0,0){
     \setlength{\unitlength}{.75pt}\put(-34,-15){
     \put(60, 67) {\scriptsize$ (1{,}2) $}
     \put(60, 31) {\scriptsize$ (1{,}2) $}
     \put(128, 62) {\scriptsize$ b_{\lambda,\mu} $}
     \put( 95, 61) {\scriptsize$ \phi^2 $}
     }\setlength{\unitlength}{1pt}}
  \end{picture}}
\\[3em]\displaystyle
\hspace{4em} = ~
\mu \,
  \raisebox{-25pt}{
  \begin{picture}(100,58)
   \put(0,0){\scalebox{.75}{\includegraphics{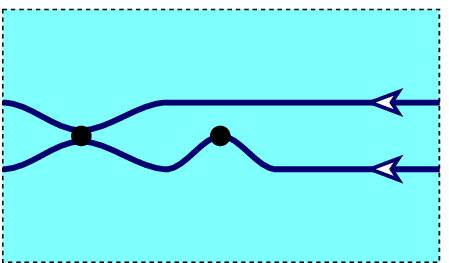}}}
   \put(0,0){
     \setlength{\unitlength}{.75pt}\put(-34,-15){
     \put(120, 67) {\scriptsize$ (1{,}2) $}
     \put(120, 31) {\scriptsize$ (1{,}2) $}
     \put( 45, 62) {\scriptsize$ b_{\lambda,\mu} $}
     \put( 92, 37) {\scriptsize$ \phi^2 $}
     }\setlength{\unitlength}{1pt}}
  \end{picture}}
~+~
\lambda \,
  \raisebox{-25pt}{
  \begin{picture}(100,58)
   \put(0,0){\scalebox{.75}{\includegraphics{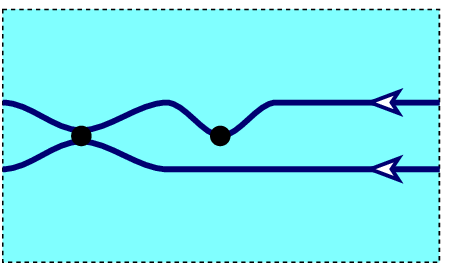}}}
   \put(0,0){
     \setlength{\unitlength}{.75pt}\put(-34,-15){
     \put(120, 67) {\scriptsize$ (1{,}2) $}
     \put(120, 31) {\scriptsize$ (1{,}2) $}
     \put( 45, 62) {\scriptsize$ b_{\lambda,\mu} $}
     \put( 95, 61) {\scriptsize$ \phi^2 $}
     }\setlength{\unitlength}{1pt}}
  \end{picture}}
  \quad .
\eear\ee
To prove this, one uses the orthonormality and completeness relations \eqref{eq:fusion-orthonormality} and  \eqref{eq:fusion-completeness}, as well as the bubble-collapse identities \eqref{eq:bubble-collapse}. It is enough to verify 
$\jun^{d \rightarrow 2\star 2} (\text{l.h.s}) \bar\jun^{2\star 2 \rightarrow c} = \jun^{d \rightarrow 2\star 2} (\text{r.h.s}) \bar\jun^{2\star 2 \rightarrow c}$ for $c,d \in \{1,3\}$. For example, in the case $c=1$, $d=1$ one gets zero on both sides (there is no defect field of weight $(0,h_3)$ on the identity defect), and for $c=3$, $d=1$ one finds 
\bea
\jun^{1 \rightarrow 2\star 2}  \, 
b_{\lambda,\mu} \,
  \big(\lambda (\phi^2 {\times}\one^2)(x) + \mu (\one^2{\times}\phi^2)(x)\big) \,
\bar\jun^{2\star 2 \rightarrow 3}
\enl
\quad = (\mu q^{-1} - \lambda q)\, 
\big( \lambda + \tfrac{\Rs^{(22)1}}{\Rs^{(22)3}} \mu \big) \frac{\eta^{22}}{\eta^{31}}
\Gs^{(322)1}_{23} \cdot  \phi^{1 \rightarrow 3}(x)
\eear\labl{eq:b-field-aux1}
and
\bea
\jun^{1 \rightarrow 2\star 2}  \, 
  \big(\mu (\phi^2{\times}\one^2)(x) + \lambda (\one^2{\times}\phi^2)(x)\big) \, 
  b_{\lambda,\mu} \,
\bar\jun^{2\star 2 \rightarrow 3}
\enl
\quad = (\lambda q^{-1} - \mu q)\, 
\big( \mu + \tfrac{\Rs^{(22)1}}{\Rs^{(22)3}} \lambda \big) \frac{\eta^{22}}{\eta^{31}}
\Gs^{(322)1}_{23} \cdot  \phi^{1 \rightarrow 3}(x) \ .
\eear\labl{eq:b-field-aux2}
From \eqref{eq:mm-Rs-val} one computes 
$\Rs^{(22)1}/\Rs^{(22)3} = - q^2$, and substituting this shows that
\eqref{eq:b-field-aux1} and \eqref{eq:b-field-aux2} agree.
The remaining cases for the first identity in \eqref{eq:b-exchange-relations}, as well as the second identity in \eqref{eq:b-exchange-relations} can be established in the same way.
For the composition of two $b$'s one finds
\be
  b_{\lambda,\mu} b_{\mu,\lambda} = \varphi(\lambda,\mu) \cdot \one^{2 \star 2} 
  ~~ , \quad \text{with} ~~
  \varphi(\lambda,\mu) = (\lambda q^{-1}-\mu q)(\mu q^{-1}-\lambda q)  \ .
\labl{eq:bb-composition}
This implies that $b_{\lambda,\mu}$ has an inverse, provided that $\mu \neq \lambda q^{\pm 2}$.

In the next section we will perturb the $(1,s)$-defect by the defect field
\be
  \psi^s_{\lambda,\tilde\lambda}(z) = 
  \lambda \phi^s(z) + \tilde\lambda \bar\phi^s(z) \ .
\labl{eq:psi-ll-def}
Instead of the separate identities \eqref{eq:b-exchange-relations}, we will actually require a combined identity for $\psi^2_{\lambda,\tilde\lambda}$ and $\psi^2_{\mu,\tilde\mu}$, i.e.\ we would like
\be
  b_{\lambda,\mu} \,
  \big( (\psi^2_{\lambda,\tilde\lambda}{\times}\one^2)(x) + 
  (\one^2{\times} \psi^2_{\mu,\tilde\mu}(x) \big)
\overset{?}{=}  
  \big( (\psi^2_{\mu,\tilde\mu}{\times}\one^2)(x) + 
  (\one^2{\times} \psi^2_{\lambda,\tilde\lambda}(x) \big) \,
  b_{\lambda,\mu}  \ .
\labl{eq:b-psi-exchange-relation}
This would follow from \eqref{eq:b-exchange-relations} if $b_{\lambda,\mu} = C \, b_{\tilde\mu,\tilde\lambda}$ for some $C \in \Cb$. From the explicit expression \eqref{eq:b-def} it is easy to see that if
$\lambda\tilde\lambda = \mu\tilde\mu$, then
$\tilde\lambda b_{\lambda,\mu} = \mu b_{\tilde\mu,\tilde\lambda}$
and
$\tilde\mu b_{\lambda,\mu} = \lambda b_{\tilde\mu,\tilde\lambda}$.
Thus
\be
 \text{Eqn.\ \eqref{eq:b-psi-exchange-relation} holds if $\lambda\tilde\lambda = \mu\tilde\mu$
 and either $\tilde\lambda \neq 0$ or $\tilde\mu \neq 0$}\ .
\ee

\sect{Perturbed defect operators}\label{sec:pert-def}

\subsection{Definition of perturbed defects}

Let $D_s(\psi)$ denote the operator obtained by placing the $(1,s)$-defect on the unit circle and perturbing it by the defect field $\psi$. In more detail, we set
\be
  D_s(\psi) = \sum_{n=0}^\infty \frac{1}{n!} D_s^{(n)}(\psi)
  ~~, \quad
  D_s^{(n)}(\psi) = \int_0^{2\pi} \hspace{-.5em}
  D_s[\psi(\theta_1),\dots,\psi(\theta_n)] \,
  d\theta_1 \cdots d\theta_n \ ,
\ee
where the notation $D_s[\cdots]$ was introduced in \eqref{eq:defect-with-fields}. We will be interested in the fields $\psi^s_{\lambda,\tilde\lambda}$ defined in \eqref{eq:psi-ll-def}. The relation to \eqref{eq:Ds-lamtlam} in the introduction is
\be
  D_s(\lambda,\tilde\lambda) \equiv D_s(\psi^s_{\lambda,\tilde\lambda}) \ .
\ee
The OPE $\psi^s_{\lambda,\tilde\lambda}(x)\psi^s_{\lambda,\tilde\lambda}(y)$ does not have short distance singularities (recall our choice $t<\frac12$ in \eqref{eq:t-def}), and so the individual multiple integrals $D_s^{(n)}(\psi^s_{\lambda,\tilde\lambda})$ are well-defined. It is less clear if or in what sense the infinite sum in $D_s(\psi^s_{\lambda,\tilde\lambda})$ converges, but one would expect the individual matrix elements of this operator to have at least a finite radius of convergence (in $\lambda$ and $\tilde\lambda$). As mentioned in the introduction, the question of convergence will not be addressed here and the $D_s(\lambda,\tilde\lambda)$ are to be understood as formal power series in $\lambda$ and $\tilde\lambda$ with coefficients in the linear maps from $\Hc$ to the algebraic completion $\overline\Hc$. For ease of exposition, I will, however, treat $\lambda$ and $\tilde\lambda$ as complex numbers.

The first few terms of $D_s(\psi^s_{\lambda,\tilde\lambda})$ read
\bea
  D_s(\psi^s_{\lambda,\tilde\lambda})
  = D_s 
  + \int_0^{2\pi} \hspace{-.3em} \Big( 
     \lambda D_s[\phi^s(\theta)] + \tilde\lambda D_s[\bar\phi^s(\theta)]  \Big) d\theta
\enl
  + \int_0^{2\pi} \hspace{-.3em} \Big( 
     \tfrac12 \lambda^2 D_s[\phi^s(\theta_1),\phi^s(\theta_2)] 
     + \lambda \tilde\lambda D_s[\phi^s(\theta_1),\bar\phi^s(\theta_2)] 
     + \tfrac12 \tilde\lambda^2 D_s[\bar\phi^s(\theta_1),\bar\phi^s(\theta_2)] 
     \Big) d\theta_1 d\theta_2
\enl
  +~ O(\lambda^m \tilde\lambda^n; m{+}n{=}3) \ .
\eear\ee
As an example, consider the matrix element $\langle 0 | D_s(\psi^s_{\lambda,\tilde\lambda}) | \Phi\rangle$ to this order. The only term which does not involve a vanishing chiral one-point block in the holomorphic or anti-holo\-morphic factor of the correlator is the $\lambda \tilde\lambda$-term. Thus,
\be
  \langle 0 | D_s(\psi^s_{\lambda,\tilde\lambda}) | \Phi\rangle
  = \lambda \tilde\lambda \cdot E + O(\lambda^m \tilde\lambda^n; m{+}n{=}3) 
\labl{eq:0-Phi-element-lowest-order}
with $E = \int_0^{2\pi}
  \langle 0|\,D_s[\phi^s(\theta_1),\bar\phi^s(\theta_2)] \, | \Phi\rangle
   \, d\theta_1 d\theta_2$.
The correlator in the integrand has been computed in Section~\ref{sec:1bulk1def-ex}, and inserting the result we find
\be
  E =  2 \pi C_s C_{\Phi\Phi}^{~\one} 
  \langle \one \rangle \int_0^{2\pi} \hspace{-0.7em} e^{- i h_3 \delta} d\delta
  = \big(\eta^{(s)}_\phi\big)^2 \, \eta_\Phi \,  \langle \one \rangle \cdot 
  \frac{-4 \pi \sin(2 \pi t)}{h_3} \, \dim(s) \, \Fs^{(33s)s}_{s1} \ .
\ee

\subsection{Composition rules}

The composition of two unperturbed defect operators $D_a$ and $D_b$, for $a,b \in \Ic$ just gives the defect operator for the fused defect $D_{a \star b}$, which can be further decomposed according to the fusion rules
\be
  D_a D_b = D_{a \star b} = \sum_{c \in \Ic} N_{ab}^{~c} \, D_c \ .
\ee
This can be verified using \eqref{eq:Da-operator} and the Verlinde formula \cite{Verlinde:1988sn}. For perturbed defects we still have
\be
  D_a(\psi) D_b(\phi) = D_{a \star b}( \psi{\times}\one^b + \one^a{\times}\phi ) \ ,
\ee
but the right hand side generally does no longer decompose into a direct sum (see \cite[Sect.\,2.4]{Runkel:2007wd} and \cite[Sect.\,3.3]{Manolopoulos:2009np}). Let $d = \bigoplus_i c_i$ be the direct sum decomposition of $a \star b$, i.e.\ the set $\{c_i\}$ is given by all $c_i$ for which $N_{ab}^{~c_i} =1$. We can write
\be
  D_{a \star b}( \psi{\times}\one^b + \one^a{\times}\phi ) 
  = D_d( {\textstyle \sum_{i,j}} \, \xi^{c_i\rightarrow c_j} )
   ~~ , \quad
     \xi^{c_i\rightarrow c_j} 
     = \jun^{c_i\rightarrow a\star b} (\psi{\times}\one^b + \one^a{\times}\phi )
    \bar\jun^{a \star b \rightarrow c_j} \ .
\ee
The reason that $D_d$ can in general not be written as a sum of other defect operators is that the perturbation $\xi^{c_i\rightarrow c_j}$ joins the different defects in the superposition $d = \bigoplus_i c_i$. If, however, enough of the $\xi^{c_i\rightarrow c_j}$ are zero, we may still obtain a sum decomposition. This is what happens in the example we are considering in the present paper. 

Suppose the defect fields $\phi^{(s)}$ and $\bar\phi^{(s)}$ are normalised via \eqref{eq:phi-phibar-OPE} and \eqref{eq:defect-structure-constants}, and that the normalisation constants $\eta^{(s)}_\phi$ are chosen in a correlated fashion as in \eqref{eq:eta-s-relation}. 
For $s \in \{ 2, \dots, p'-2 \}$ consider the composition
\bea
  D_2(\psi_{\lambda,\tilde\lambda}) D_s(\psi_{\mu,\tilde\mu})
  =
  D_{2 \star s}(
  \lambda \phi^2{\times}\one^s + \mu \one^2{\times}\phi^s +
  \tilde\lambda \bar\phi^2{\times}\one^s + \tilde\mu \one^2{\times}\bar\phi^s)
\enl
  = D_{(s{-}1)\oplus(s{+}1)}( {\textstyle \sum_{\sigma,\nu = \pm 1}}~
  \xi_{\sigma,\nu} \cdot \phi^{s+\nu \rightarrow s+\sigma} + 
   \bar\xi_{\sigma,\nu}\cdot \bar\phi^{s+\nu \rightarrow s+\sigma})
\eear\ee
where
\bea
   \xi_{\sigma,\nu} \cdot \phi^{s+\nu \rightarrow s+\sigma}
   = \jun^{s{+}\nu\rightarrow 2\star s} 
   (\lambda \phi^2{\times}\one^s + \mu \one^2{\times}\phi^s)
    \bar\jun^{2 \star s \rightarrow s{+}\sigma} 
\enl
   \bar\xi_{\sigma,\nu}\cdot \bar\phi^{s+\nu \rightarrow s+\sigma}
   = \jun^{s{+}\nu\rightarrow 2\star s} 
   (\tilde\lambda \bar\phi^2{\times}\one^s + \tilde\mu \one^2{\times}\bar\phi^s)
    \bar\jun^{2 \star s \rightarrow s{+}\sigma} 
\eear\ee
Substituting \eqref{eq:bubble-collapse} one finds that
\bea
  \xi_{\nu,\nu} = 
  \frac{\sin(\pi t)}{\sin(\pi \nu s t)} \, \lambda + 
  \frac{\sin(\pi(\nu s{-}1)t)}{\sin(\pi \nu s t)} \, \mu \ ,
\enl
  \bar\xi_{\nu,\nu} = 
  \frac{\sin(\pi t)}{\sin(\pi \nu s t)} \, \tilde\lambda + 
  \frac{\sin(\pi(\nu s{-}1)t)}{\sin(\pi \nu s t)} \, \tilde\mu \ ,
\enl
  \xi_{\nu,-\nu} = 
  \frac{\eta^{22}}{\eta^{s+\nu,s-\nu}} \,
  \frac{ \Ga{2{-}3t} \Ga{\nu(1{-}st)} }{ \Ga{1{-}2t} \Ga{1{+}\nu{-}(\nu s{+}1)t} }
  \, \big( \lambda - q^{\nu s} \mu \big) \ ,
\enl
  \bar\xi_{\nu,-\nu} =
  \frac{\eta^{22}}{\eta^{s+\nu,s-\nu}} \,
  \frac{ \Ga{2{-}3t} \Ga{\nu(1{-}st)} }{ \Ga{1{-}2t} \Ga{1{+}\nu{-}(\nu s{+}1)t} }
  \, \big( \tilde\mu - q^{\nu s} \tilde\lambda \big) \ .
\eear\ee
If we choose
\be
  \mu = q^{s\eps} \lambda
\quad \text{and}\quad
  \tilde\mu = q^{-s\eps} \tilde\lambda
\ee  
for $\eps \in \{ \pm 1 \}$, then $\xi_{-\eps,\eps}=\bar\xi_{-\eps,\eps}=0$
and $\xi_{\nu,\nu} = q^{ \eps(s-\nu ) } \lambda$, 
$\bar\xi_{\nu,\nu} = q^{ -\eps(s-\nu ) } \tilde\lambda$.
For either choice of $\eps$, the field which changes the defect condition is missing its partner and, as explained in \cite[Sect.\,2.4]{Runkel:2007wd} and \cite[Sect.\,3.3]{Manolopoulos:2009np}, it then cannot contribute to the perturbed defect operator. This establishes the identity \eqref{eq:D2Ds-composition} from the introduction.

The defect $D_{p'-1}$ obeys $D_{p'-1} D_{p'-1} = D_1 = \id$ and implements the $\Zb_2$-symmetry of $M(p,p')$ (it is a group-like defect in the sense of \cite{Frohlich:2006ch}). If we fuse it with a perturbed defect, we obtain the identity
\be
  D_{p'-1} D_s( \psi_{\lambda,\tilde\lambda} ) = 
  D_{p'-s}( \lambda \xi + \tilde\lambda \bar\xi ) \ ,
\ee
where, using \eqref{eq:eta-s-relation}, \eqref{eq:bubble-collapse} and \eqref{eq:F(ps3)}, one obtains \cite{Runkel:2007wd}
\be
  \xi = \jun^{p'{-}s\rightarrow (p'{-}1)\star s} ( \one^{p'-1}{\times}\phi^s ) 
      \bar\jun^{(p'{-}1)\star s \rightarrow p'{-}s} 
      = \frac{\eta^{(s)}_\phi}{\eta^{(p'-s)}_\phi} \Fs^{(p'-1,s,3)p'-s}_{s,p'-s} \phi^{p'-s}
      = (-1)^p \phi^{p'-s} \ ,
\ee
and similarly $\bar\xi = \jun^{p'{-}s\rightarrow (p'{-}1)\star s} ( \one^{p'-1}{\times}\bar\phi^s ) 
      \bar\jun^{(p'{-}1)\star s \rightarrow p'{-}s} = (-1)^p \, \bar\phi^{p'-s}$.
One checks that $D_s( \psi_{\lambda,\tilde\lambda} ) D_{p'-1}$ gives the same result, and altogether this establishes \eqref{eq:Dinv-Ds} from the introduction.

\subsection{Defect operators mutually commute}

In this section we will prove the commutator \eqref{eq:DmDn-comm}, i.e.\ that $[D_m(\lambda,\tilde\lambda),D_n(\mu,\tilde\mu)] = 0$ for $\lambda \tilde\lambda = \mu \tilde\mu$. Let us from now on assume that $\lambda \tilde\lambda = \mu \tilde\mu$. We will distinguish three cases. 

\subsubsection*{Case 1: $\tilde\lambda = 0$ and $\tilde\mu = 0$}

This case was already considered in \cite{Runkel:2007wd}, and the commutator \eqref{eq:DmDn-comm} was established in \cite[Sect.\,3.1]{Runkel:2007wd}. 

\subsubsection*{Case 2: ($\tilde\lambda \neq 0$ or $\tilde\mu \neq 0$) and $\lambda = 0$ and $\mu = 0$}

The proof of case 2 works along the same lines as that of case 1, and we skip the details.

\subsubsection*{Case 3: ($\tilde\lambda \neq 0$ or $\tilde\mu \neq 0$) and ($\lambda \neq 0$ or $\mu \neq 0$)}

The functional relation \eqref{eq:D2Ds-composition} allows to express $D_s(\lambda,\tilde\lambda)$ as a polynomial in $D_2(\rho,\tilde\rho)$, where the parameters $\rho$, $\tilde\rho$ differ from $\lambda$, $\tilde\lambda$ by phases which may vary from term to term, but they always satisfy $\rho\tilde\rho = \lambda\tilde\lambda$. To establish $[D_m(\lambda,\tilde\lambda),D_n(\mu,\tilde\mu)] = 0$, it is therefore enough to show
\be
  [D_2(\lambda,\tilde\lambda),D_2(\mu,\tilde\mu)] = 0   
\labl{eq:D2D2-case3}
for all $\lambda$, $\tilde\lambda$, $\mu$, $\tilde\mu$ which obey $\lambda \tilde\lambda = \mu \tilde\mu$. Since case 1 and 2 are already proved, it is remains to verify \eqref{eq:D2D2-case3} under the condition of case 3, i.e.\ assuming that one of $\tilde\lambda,\tilde\mu$ is non-zero and that one of $\lambda,\mu$ is non-zero.

\medskip\noindent
\nxt Suppose that for a given $\eps = \pm 1$ we have $\mu = q^{2 \eps} \lambda$. By assumption, $\lambda$ and $\mu$ cannot both be zero, and hence they are both non-zero. By $\lambda \tilde\lambda = \mu \tilde\mu$ and the assumption that one of $\tilde\lambda,\tilde\mu$ is non-zero, this implies that all four constants $\lambda$, $\tilde\lambda$, $\mu$, $\tilde\mu$ are non-zero, and that $\tilde\mu = q^{- 2 \eps} \tilde\lambda$. The composition rule \eqref{eq:D2Ds-composition} applies and gives
\be
  D_2(\lambda,\tilde\lambda) \, D_2(\mu,\tilde\mu) = 
  D_2(\lambda,\tilde\lambda) \,
  D_2(q^{2\eps} \lambda, q^{-2\eps} \tilde\lambda) 
  ~=~
  D_{1} 
  ~+~ 
  D_{3}(q^{\eps} \lambda, q^{-\eps} \tilde\lambda)  \ .
\ee
The composition in the opposite order can also be computed with 
\eqref{eq:D2Ds-composition} and gives
\be
   D_2(\mu,\tilde\mu) \, D_2(\lambda,\tilde\lambda)  = 
  D_2(q^{2\eps} \lambda, q^{-2\eps} \tilde\lambda) \,
  D_2(\lambda,\tilde\lambda) 
  ~=~
  D_{1} 
  ~+~ 
  D_{3}(q^{\eps} \lambda, q^{-\eps} \tilde\lambda)  \ .
\ee
Thus $[D_2(\lambda,\tilde\lambda),D_2(\mu,\tilde\mu)] = 0$.

\medskip\noindent
\nxt Suppose that $\lambda \neq q^{\pm 2} \mu$. Then according to \eqref{eq:bb-composition} the operator $b_{\lambda,\mu}$ is invertible, and by the assumptions for case 3, the identity \eqref{eq:b-psi-exchange-relation} holds. Set
\be
  \psi = \psi^2_{\lambda,\tilde\lambda} {\times} \one^2 + \one^2 {\times}  \psi^2_{\mu,\tilde\mu} 
  \quad , \quad
  \psi' = \psi^2_{\mu,\tilde\mu} {\times} \one^2 + \one^2 {\times}  \psi^2_{\lambda,\tilde\lambda} 
  \ .
\ee
Then
\be
  D_2(\lambda,\tilde\lambda) \, D_2(\mu,\tilde\mu) = D_{2 \star 2}(\psi)
  \quad , \quad
  D_2(\mu,\tilde\mu) \, D_2(\lambda,\tilde\lambda)  = D_{2 \star 2}(\psi') \ ,
\ee
and we need to show that $D_{2 \star 2}(\psi) = D_{2 \star 2}(\psi')$. 

Let us call an ordered list of mutually distinct angles $( \theta_1, \dots ,\theta_n )$ (which we identify with points on the unit circle) {\em cyclically ordered} iff for each $k \in \{1,\dots,n\}$, starting at $\theta_k$ and moving counter-clockwise along the unit circle, the next value one meets is $\theta_{k+1}$ (or $\theta_1$ if $k=n$). Given $n$ mutually distinct points $( \theta_1, \dots ,\theta_n )$ on the unit circle there is a permutation $\sigma \in S_n$ (unique up to shifts)  such that $( \theta_{\sigma(1)}, \dots, \theta_{\sigma(n)} )$ is cyclically ordered.

Consider the integrand of
$D^{(n)}_{2 \star 2}(\psi)$ for a choice of mutually distinct $\theta_1,\dots,\theta_n$ (if two angles are equal, the integrand is zero). Pick a permutation $\sigma \in S_n$ and angles $\alpha_1,\dots,\alpha_n$ such that
\be
  ( \theta_{\sigma(1)}, \alpha_1, \theta_{\sigma(2)} , \alpha_2 ,  \dots, \theta_{\sigma(n)}, \alpha_n )
\ee
is a cyclically ordered set of mutually distinct angles. Then
\bea
  D_{2\star 2}[  \psi(\theta_{\sigma(1)}), \dots, \psi(\theta_{\sigma(n)}) ]
\enl
\quad  \overset{(1)}{=}~ 
    \varphi(\lambda,\mu)^{-1}  
    D_{2\star 2}[  \psi(\theta_{\sigma(1)}), \dots, \psi(\theta_{\sigma(n)}), (b_{\lambda,\mu} b_{\mu,\lambda})(\alpha_{n}) ]
\enl
\quad  \overset{(2)}{=}~ 
    \varphi(\lambda,\mu)^{-1}  
    D_{2\star 2}[  \psi(\theta_{\sigma(1)}), \dots, b_{\lambda,\mu}(\alpha_{n-1}), \psi'(\theta_{\sigma(n)}),  b_{\mu,\lambda}(\alpha_{n}) ]
\enl
\quad  =~ 
  \cdots
\enl
\quad  \overset{(3)}{=}~ 
    \varphi(\lambda,\mu)^{-1}  
    D_{2\star 2}[  \psi'(\theta_{\sigma(1)}), \dots, \psi'(\theta_{\sigma(n)}),  (b_{\mu,\lambda}b_{\lambda,\mu})(\alpha_{n}) ]
\enl
\quad  \overset{(4)}{=}~ 
  D_{2\star 2}[  \psi'(\theta_{\sigma(1)}), \dots, \psi'(\theta_{\sigma(n)}) ] \ ,
\eear\ee
where in step 1 we inserted the identity $\one^{2 \star 2}$ according to \eqref{eq:bb-composition}; $\varphi(\lambda,\mu)^{-1}$ exists as by assumption $\lambda \neq q^{\pm 2} \mu$. In step 2 we used \eqref{eq:b-psi-exchange-relation} once, namely $b_{\lambda,\mu} \psi = \psi' b_{\lambda,\mu}$. Repeating this procedure brings us to equality 3, and step 4 amounts to \eqref{eq:bb-composition} and $\varphi(\lambda,\mu)=\varphi(\mu,\lambda)$. Altogether, this shows that $D^{(n)}_{2 \star 2}(\psi) = D^{(n)}_{2 \star 2}(\psi')$ for all $n$ and thus implies \eqref{eq:D2D2-case3} also in the case $\lambda \neq q^{ \pm 2} \mu$.

\medskip

We have now established the commutator \eqref{eq:DmDn-comm} in the introduction for all values of $m,n \in \{2,\dots,p'{-}2\}$ and $\lambda, \tilde\lambda, \mu, \tilde\mu \in \Cb$ with $\lambda \tilde\lambda = \mu \tilde\mu$.

\subsection{Commutator with perturbed Hamiltonian}

Recall the definition of $\phi\bar\phi^s$ and $\bar\phi\phi^s$ from \eqref{eq:phiphibar-def}. Introduce the `commutator' field
\be
  \Delta^{\!s}(x) = \phi\bar\phi^s(x) - \bar\phi\phi^s(x) \ .
\ee
on the $(1,s)$-defect. The following identity involving the $(1,2)$-defect is the key observation to understand the commutator of the perturbed defect and the perturbed Hamiltonian,
\be
  \raisebox{-25pt}{
  \begin{picture}(60,58)
   \put(0,0){\scalebox{.75}{\includegraphics{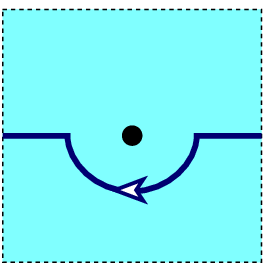}}}
   \put(0,0){
     \setlength{\unitlength}{.75pt}\put(-34,-15){
     \put (67, 59) {\scriptsize$ \Phi(x) $}
     \put (78, 26) {\scriptsize$ (1{,}2) $}
     }\setlength{\unitlength}{1pt}}
  \end{picture}}
~-~
  \raisebox{-25pt}{
  \begin{picture}(60,58)
   \put(0,0){\scalebox{.75}{\includegraphics{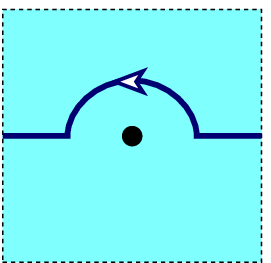}}}
   \put(0,0){
     \setlength{\unitlength}{.75pt}\put(-34,-15){
     \put (67, 38) {\scriptsize$ \Phi(x) $}
     \put (78, 73) {\scriptsize$ (1{,}2) $}
     }\setlength{\unitlength}{1pt}}
  \end{picture}}
~=~ C_{[\Phi]} \,
  \raisebox{-25pt}{
  \begin{picture}(60,58)
   \put(0,0){\scalebox{.75}{\includegraphics{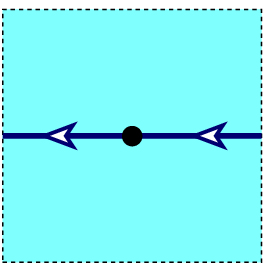}}}
   \put(0,0){
     \setlength{\unitlength}{.75pt}\put(-34,-15){
     \put (62, 59) {\scriptsize$ \Delta^2(x) $}
     \put (82, 38) {\scriptsize$ (1{,}2) $}
     }\setlength{\unitlength}{1pt}}
  \end{picture}}
\labl{eq:Phi-D2-comm}
where
\be
  C_{[\Phi]} = \frac{\eta_\Phi}{\big(\eta^{(2)}_\phi\big)^2} \cdot (-2i) \sin(3\pi t) \,
  \frac{\Ga{2{-}2t}\Ga{1{-}2t}}{\Ga{2{-}3t}\Ga{1{-}t}} \ .
\labl{eq:C-Phi-comm-const}
This identity will be derived in Section~\ref{sec:TFT-struct} using the TFT approach. With the help of \eqref{eq:Phi-D2-comm} we can compute, for $\psi \equiv \psi_{\lambda,\tilde\lambda}$,
\bea
  \Big[ ~D_2(\psi) ~,~ {\textstyle \int_0^{2\pi}} \Phi(e^{i \alpha}) d\alpha ~\Big]
\enl
  \qquad =~ 
  \sum_{n=0}^\infty \frac{1}{n!}
  \int_0^{2\pi} 
  \Big[~  D_2[ \psi(\theta_1), \dots ,\psi(\theta_n)] ~,~ \Phi(e^{i \alpha})  ~\Big] 
  d\alpha \, d\theta_1 \cdots d\theta_n
\enl
  \qquad =~ 
  C_{[\Phi]} \sum_{n=0}^\infty \frac{1}{n!}
  \int_0^{2\pi} \hspace{-.7em}
  D_2[ \psi(\theta_1), \dots ,\psi(\theta_n),\Delta^{\!2}(\alpha) ] \,
  d\alpha \, d\theta_1 \cdots d\theta_n
\eear\labl{eq:pert-Ham-comm-aux4}
Next we analyse the commutator of $L_0{+}\bar L_0$ with $D_2(\psi)$. On the $(1,2)$-defect define the field $\omega(x)$ as
\be
  \omega(x) = \lambda \,\phi^2(x) - \tilde\lambda \,\bar\phi^2(x) \ .
\ee
Note that $[L_0{+}\bar L_0,\phi^2(x)] = [L_0,\phi^2(x)] = [L_0{-}\bar L_0,\phi^2(x)]$, and that similarly $[L_0{+}\bar L_0,\bar\phi^2(x)] = - [L_0{-}\bar L_0,\bar\phi^2(x)]$. Consider a $(1,2)$-defect placed on the unit circle with clockwise orientation, and with an insertion of $\psi$ at the point $e^{i\theta}$. Since $i(L_0-\bar L_0)$ is the generator of rotations, with our choice of local coordinates we have
\be
  \big[ L_0+\bar L_0 , \psi(e^{i\theta}) \big] =   \big[ L_0-\bar L_0 , \omega(e^{i\theta}) \big] 
  = - i \tfrac{\partial}{\partial\theta} \omega(e^{i\theta}) \ .
\ee
Thus in particular
\be
  \Big[~ L_0+\bar L_0 ~,~ D_2[ \psi(\theta_1),\dots,\psi(\theta_n)] ~\Big] = 
  -i \sum_{k=1}^n  \tfrac{\partial}{\partial\theta_k} D_2[ \psi(\theta_1),\dots, \omega(\theta_k),\dots,\psi(\theta_n)] \ .
\labl{eq:pert-Ham-comm-aux1}
Integrating over $\theta_1,\dots,\theta_n$ results in
\be
  \big[ L_0+\bar L_0 ,  D^{(n)}_2(\psi) \big] 
  =  -i\,n \int_0^{2\pi} \hspace{-.4em}
   \tfrac{\partial}{\partial\alpha} D_2[ \psi(\theta_1),\dots, \psi(\theta_{n-1}),\omega(\alpha)]   
  d\alpha\,d\theta_1\cdots d\theta_{n-1}
\labl{eq:pert-Ham-comm-aux3}
Before carrying out the $\alpha$-integration, we have to take into account that as a function of $\alpha$ the integrand is only piece-wise smooth, as the integrand may jump when $\omega(\alpha)$ is taken past the other insertions $\psi(\theta_i)$. 
For a given set of mutually distinct values $\{ \theta_1,\dots,\theta_{n-1} \}$
pick a permutation $\sigma \in S_{n-1}$ such that $(\theta_{\sigma(1)},\dots,\theta_{\sigma(n-1)})$ is cyclically ordered. Then we can integrate over $\alpha$ on each of the segments ${]}\theta_{\sigma(k)},\theta_{\sigma(k+1)}{[}$, which results in the boundary terms $\omega(\theta_{\sigma(k+1)}-\eps)-\omega(\theta_{\sigma(k)}+\eps)$, where $\eps>0$ is included to remember the correct ordering, and has to be taken to zero. This gives
\bea
   \int_0^{2\pi} \hspace{-.4em}
   \tfrac{\partial}{\partial\alpha} D_2[ \psi(\theta_1),\dots, \psi(\theta_{n-1}),\omega(\alpha)]   
  d\alpha
\enl
\qquad =~
   \int_0^{2\pi} \hspace{-.4em}
   \tfrac{\partial}{\partial\alpha} D_2[ \psi(\theta_{\sigma(1)}),\dots, \psi(\theta_{\sigma(n-1)}),\omega(\alpha)]   
  d\alpha
\enl
\qquad =~ \sum_{k=1}^{n-1} \lim_{\eps\searrow 0}
   D_2[ \psi(\theta_{\sigma(1)}),\dots, \psi(\theta_{\sigma(n-1)}),
   \omega(\theta_{\sigma(k+1)}-\eps)]   
\enl
\qquad \qquad \qquad    -
   \sum_{k=1}^{n-1} \lim_{\eps\searrow 0} D_2[ \psi(\theta_{\sigma(1)}),\dots, \psi(\theta_{\sigma(n-1)}),
   \omega(\theta_{\sigma(k)}+\eps)]  
\eear
\labl{eq:pert-Ham-comm-aux2}
Since the OPEs of $\phi^2$ and $\bar\phi^2$ with themselves vanish for coinciding points, one finds
\be
  \lim_{\eps\rightarrow 0}\big(~ \psi(x) \omega(x-\eps) - \omega(x+\eps) \psi(x) ~\big) = - 2 \lambda \tilde\lambda \cdot \Delta^{\!2}(x) \ .
\ee
Shifting the second summation index in \eqref{eq:pert-Ham-comm-aux2} and inserting the above identity results in
\be
  \text{r.h.s. of \eqref{eq:pert-Ham-comm-aux2}} 
= - 2 \lambda \tilde\lambda
\sum_{i=1}^{n-1} 
   D_2[ \psi(\theta_{\sigma(1)}),\dots, \Delta^{\!2}(\theta_{\sigma(i)}),\dots,
   \psi(\theta_{\sigma(n-1)})]     \ .
\ee
Substituting this result into \eqref{eq:pert-Ham-comm-aux3} gives
\be
  \big[ L_0+\bar L_0 ,  D^{(n)}_2(\psi) \big] 
  =  2 i \lambda \tilde\lambda \, n(n{-}1) \int_0^{2\pi} \hspace{-.5em}
  D_2[ \psi(\theta_1),\dots, \psi(\theta_{n-2}),\Delta^{\!2}(\gamma)]   
  d\gamma\,d\theta_1\cdots d\theta_{n-2} \ .
\ee
Comparing this to \eqref{eq:pert-Ham-comm-aux4} shows that
\be
  \Big[ L_0+\bar L_0 ,  D_2(\psi) \Big] + \frac{ 2 i \lambda \tilde\lambda }{C_{[\Phi]}}
  \Big[  {\textstyle \int_0^{2\pi}} \Phi(e^{i \alpha}) d\alpha , D_2(\psi) \Big] = 0 \ ,
\ee
which is nothing but property \eqref{eq:HP-commutator} from the introduction in the
case $s=2$. For general $s$, the commutator \eqref{eq:HP-commutator} is implied
by the functional relation \eqref{eq:D2Ds-composition}, which allows to express $D_s(\lambda,\tilde\lambda)$ as a polynomial in $D_2(\gamma,\tilde\gamma)$ with $\gamma\tilde\gamma = \lambda\tilde\lambda$.

\medskip

Let us give an example of this commutator identity by considering the matrix
element $\langle 0 | \cdots |\Phi\rangle$ to lowest non-vanishing order. For the
first commutator one finds
\bea
  \langle 0 | \big[ L_0+\bar L_0 ,  D_2(\psi) \big] |\Phi\rangle
  = - \langle 0 | D_2(\psi) (L_0+\bar L_0)|\Phi\rangle
  = - 2 h_3 \langle 0 | D_2(\psi)|\Phi\rangle
\enl
  \qquad = 
  \big(\eta^{(2)}_\phi\big)^2 \, \eta_\Phi \,  \langle \one \rangle \cdot 
  8 \pi \sin(2 \pi t) \, \dim(2) \, \Fs^{(332)2}_{21} \cdot \lambda \tilde\lambda
  + O(\lambda^m \tilde\lambda^n; m{+}n{=}3) 
\enl
  \qquad = 
   \big(\eta^{(2)}_\phi\big)^2 \, \eta_\Phi \,  \langle \one \rangle \cdot 
  8 \pi \sin(2 \pi t) \, \frac{\Ga{2{-}3t}\Ga{1{-}t}}{\Ga{2{-}2t}\Ga{1{-}2t}}
  \cdot \lambda \tilde\lambda
  + O(\lambda^m \tilde\lambda^n; m{+}n{=}3) \ , 
\eear\ee
where we used \eqref{eq:0-Phi-element-lowest-order}. For the second commutator, to the given order it is enough to consider the unperturbed defects and we can use \eqref{eq:Da-operator},
\bea
  \langle 0 | \big[  {\textstyle \int_0^{2\pi}} \Phi(e^{i \alpha}) d\alpha , D_2(\psi) \big] |\Phi\rangle
  = 
  2 \pi \langle 0 | \Phi(1) D_2 |\Phi\rangle
  - 2 \pi \langle 0 | D_2 \Phi(1)  |\Phi\rangle
\enl
  \qquad
  =~ 2 \pi C_{\Phi\Phi}^{~\one} \langle\one\rangle 
  \Big( \frac{S_{2,3}}{S_{1,3}} - \frac{S_{2,1}}{S_{1,1}} \Big)
  = \big(\eta_\Phi\big)^2 \langle\one\rangle \cdot  8 \pi \sin(2\pi t) \sin(3 \pi t) \ .
\eear\ee
Substituting the explicit value for $\xi$ in \eqref{eq:HP-commutator-xi}, we conclude that indeed
$ \langle 0 | [L_0+\bar L_0 + \xi \lambda \tilde\lambda \int_0^{2\pi} \Phi(e^{i\alpha}) d\alpha,D_2(\psi)] |\Phi\rangle = 0 + O(\lambda^m \tilde\lambda^n; m{+}n{=}3)$.

\subsection{Defects acting on boundaries}

Consider the upper half plane with Cardy boundary condition labelled by $(1,1) \in \Ic$ placed on the real line. If we fuse this boundary with a topological defect labelled by $(r,s)$ we obtain the Cardy boundary condition labelled $(r,s)$ \cite{Petkova:2001ag,Frohlich:2006ch}. The fusion of defects and boundary conditions has been employed to understand relations between boundary flows \cite{Graham:2003nc}, or to obtain families of boundary flows from a defect flow \cite{Bachas:2004sy}.

Similar to the identities \eqref{eq:bubble-collapse} we can compute how the defect fields $\phi^s$ and $\bar\phi^s$ on the $(1,s)$-defect turn into boundary fields $\psi^s$ on the $(1,s)$-boundary condition, see Section~\ref{sec:3dTFT}:
\be
\lim_{d \searrow 0}
  \raisebox{-25pt}{
  \begin{picture}(60,58)
   \put(0,0){\scalebox{.75}{\includegraphics{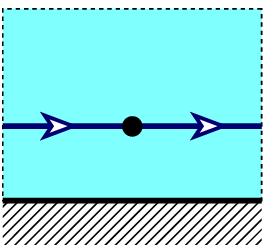}}}
   \put(0,0){
     \setlength{\unitlength}{.75pt}\put(-34,-18){
     \put (46, 36) {\scriptsize$ (1{,}1) $}
     \put (36, 59) {\scriptsize$ (1{,}s) $}
     \put (86, 59) {\scriptsize$ (1{,}s) $}
     \put (68, 60) {\scriptsize$ \chi $}
     }\setlength{\unitlength}{1pt}}
  \end{picture}}
~=~
  \raisebox{-25pt}{
  \begin{picture}(60,58)
   \put(0,0){\scalebox{.75}{\includegraphics{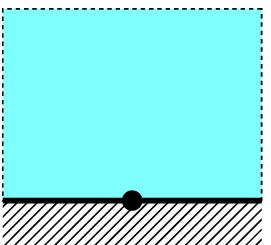}}}
   \put(0,0){
     \setlength{\unitlength}{.75pt}\put(-34,-18){
     \put (36, 36) {\scriptsize$ (1{,}s) $}
     \put (86, 36) {\scriptsize$ (1{,}s) $}
     \put (68, 40) {\scriptsize$ \psi^s $}
     }\setlength{\unitlength}{1pt}}
  \end{picture}}
\quad , \quad
\lim_{d \searrow 0}
  \raisebox{-25pt}{
  \begin{picture}(60,58)
   \put(0,0){\scalebox{.75}{\includegraphics{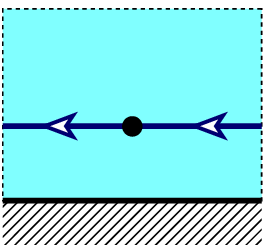}}}
   \put(0,0){
     \setlength{\unitlength}{.75pt}\put(-34,-18){
     \put (46, 36) {\scriptsize$ (1{,}1) $}
     \put (36, 59) {\scriptsize$ (1{,}s) $}
     \put (86, 59) {\scriptsize$ (1{,}s) $}
     \put (68, 60) {\scriptsize$ \chi $}
     }\setlength{\unitlength}{1pt}}
  \end{picture}}
~=~
  \raisebox{-25pt}{
  \begin{picture}(60,58)
   \put(0,0){\scalebox{.75}{\includegraphics{pic14b.eps}}}
   \put(0,0){
     \setlength{\unitlength}{.75pt}\put(-34,-18){
     \put (36, 36) {\scriptsize$ (1{,}s) $}
     \put (86, 36) {\scriptsize$ (1{,}s) $}
     \put (68, 40) {\scriptsize$ \psi^s $}
     }\setlength{\unitlength}{1pt}}
  \end{picture}}
  \quad ,
\labl{eq:bulk-bnd-bubble}
where $\chi = \phi^s$ or $\chi = \bar\phi^s$, and $d$ denotes the distance from $\chi$ to the boundary. The limit is non-singular and one could draw $\chi$ directly on the boundary, 
but this would make the picture less legible.

Denote by $|(1,s)+\gamma \psi^s\rangle\!\rangle$ the state describing the $(1,s)$-boundary condition on the complex plane minus the open unit disc, perturbed by $\gamma\cdot\psi^s$. Correspondingly,
$\langle\!\langle(1,s)+\gamma \psi^s|$ denotes the $(1,s)$-boundary condition on the closed unit disc, perturbed by $\gamma\cdot\psi^s$. By \eqref{eq:bulk-bnd-bubble} we have
\be
  D_s(\lambda,\tilde\lambda) |(1,1)\rangle\!\rangle
  = |(1,s) + (\lambda{+}\tilde\lambda) \psi^s\rangle\!\rangle  
~~ , ~~
  \langle\!\langle(1,1)| D_s(\lambda,\tilde\lambda) 
  = \langle\!\langle(1,s)+(\lambda{+}\tilde\lambda) \psi^s| \ .
\labl{eq:pert_bnd_def}
Let $Z_{s,s'}(\mu;\gamma,\gamma')$ be the partition function of a cylinder of circumference $2\pi$ and length $L$, perturbed by $\mu\cdot \Phi$, with boundary condition $(1,s)$  perturbed by $\gamma \cdot\psi^s$ on one end, and $(1,s')$ perturbed by $\gamma'\cdot \psi^{s'}$ on the other end:
\be
  Z_{s,s'}(\mu;\gamma,\gamma') =  \langle\!\langle(1,s)+ \gamma \psi^s| ~ e^{-L H(\mu)} ~
  |(1,s') + \gamma' \psi^s\rangle\!\rangle \ ,
\ee
where $H = L_0 + \bar L_0 - \tfrac{c}{12} + \mu \int_0^{2\pi} \Phi(e^{i\theta}) d\theta$. Combining \eqref{eq:tsys-relation}, \eqref{eq:HP-commutator}, \eqref{eq:pert_bnd_def} and setting $\mu = \xi \lambda \tilde\lambda$, we obtain the following identity for cylinder partition functions:
\bea
  Z_{s,s}(\mu;q \lambda+q^{-1}\tilde\lambda,q^{-1}\lambda+q\tilde\lambda) 
\enl
  =   \langle\!\langle(1,1)| D_s(q \lambda,q^{-1} \tilde\lambda) e^{-L H(\mu)}
D_s(q^{-1}\lambda,q  \tilde\lambda)  |(1,1)\rangle\!\rangle
\enl
  =   \langle\!\langle(1,1)| e^{-L H(\mu)}  |(1,1)\rangle\!\rangle
  +   \langle\!\langle(1,1)| D_{s-1}(\lambda, \tilde\lambda) e^{-L H(\mu)}
D_{s+1}(\lambda,\tilde\lambda)  |(1,1)\rangle\!\rangle
\enl
= Z_{1,1}(\mu;-,-) + Z_{s-1,s+1}(\mu;\lambda+\tilde\lambda,\lambda+\tilde\lambda) \ .
\eear\labl{eq:cyl-identity}
Choose square roots $\sqrt{\xi}$ and $\sqrt{\mu}$ for $\xi$ and $\mu$ and
parametrise $\lambda$ and $\tilde\lambda$ in terms of $\alpha \in \Cb$ as
\be
  \lambda = \frac{e^{\pi i \alpha t/2} \, \sqrt{\mu}}{i \, \sqrt{\xi}}  
  \quad , \quad
  \tilde\lambda = \frac{e^{-\pi i \alpha t/2} \, \sqrt{\mu}}{(-i) \, \sqrt{\xi}}  \ ,
\ee
so that $\xi \lambda \tilde\lambda = \mu$. Introduce the function
\be
  \gamma(\alpha) = \lambda+\tilde\lambda = \frac{2}{\sqrt{\xi}}\, \sin(\pi \alpha t/2) \cdot \sqrt{\mu} \ .
\ee
In terms of this parametrisation, the identity \eqref{eq:cyl-identity} becomes
\be
  Z_{s,s}\big(\mu;\gamma(\alpha{+}2),\gamma(\alpha{-}2)\big) 
= Z_{1,1} \big(\mu;-,-\big) + Z_{s-1,s+1} \big(\mu;\gamma(\alpha),\gamma(\alpha) \big) \ . 
\ee
This identity has been conjectured in \cite[Eqn.\,(5.32)]{Dorey:2000eh} in the case of the Lee-Yang model, where $t=2/5$, and \cite[Eqn.\,(2.22)]{Dorey:2000eh} predicts that $2/\sqrt{\xi}=-5^{3/8}/(\sqrt{2} \sin \tfrac\pi5)$. To verify this, we first have to match the normalisation convention of \cite{Dorey:2000eh}. Choose the constants $\eta_\Phi$ and $\eta^{(2)}_\phi$ such that $C_{\Phi\Phi}^{~\one}=-1$ and $C^{(2)\one}_{\phi\phi}=-1$ as in \cite[Eqn.\,(2.2)]{Dorey:2000eh}. By adapting the signs of  $\eta_\Phi$ and $\eta^{(2)}_\phi$ if necessary, one checks that also
$C_{\Phi\Phi}^{~\Phi}$ and $C^{(2)\phi}_{\phi\phi}$ agree with $C_{\varphi\varphi}^{~\varphi}$ and $C_{\phi\phi}^{~\phi}$ given there, as they have to.
Setting $t=2/5$ in \eqref{eq:HP-commutator-xi} and choosing the square root of $\xi$ as $\sqrt{\xi} = - \sqrt{|\xi|}$, one recovers \cite[Eqn.\,(2.22)]{Dorey:2000eh}.

The value for  $2/\sqrt{\xi}$ was derived in \cite{Dorey:1999cj} by combining conformal perturbation theory and the massive integrable scattering description of the perturbed Lee-Yang model \cite{Zamolodchikov:1989cf,Zamolodchikov:1995xk}. The derivation given here is based solely on conformal field theory.

\sect{Derivations using 3-dim.\ topological field theory}\label{sec:3dTFT}

This section contains an brief summary of how to use three-dimensional topological field theory as defined in \cite{Turaev:1994} to describe correlation functions of two-dimensional conformal field theory \cite{Felder:1999mq,tft1,tft4,Frohlich:2006ch}. We will restrict our attention to Virasoro minimal models and to genus zero surfaces, for this is all we will need.

\subsection{Conformal blocks, correlators and 3d\,TFT}

An extended Riemann sphere $\Sigma$ is the Riemann sphere $\hat\Cb = \Cb \cup \{ \infty \}$ with a finite ordered set of marked points $(z_i,\theta_i,V_i)$, $i=1,\dots,n$, where $z_i \in \Cb$ are mutually distinct points, $\theta_i$ are angles, and $V_i$ is a representation of the Virasoro algebra, isomorphic to a finite direct sum of irreducible representations $R_a$, $a \in \Ic$. The angle $\theta_i$ defines a local coordinate $f_{z_i,\theta_i}$ around $z_i$ as given in \eqref{eq:local-coord-theta}. We write
\be
  \Sigma = \big( (z_1,\theta_1,V_1),\dots,(z_n,\theta_n,V_n) \big) \ .
\labl{eq:Sigma-for-n-points}
To an extended Riemann sphere $\Sigma$ one assigns the space of conformal blocks
$\mathrm{Bl}(\Sigma)$, which consists of multi-linear maps $V_1 \times \cdots \times V_n \rightarrow \Cb$ compatible with the Virasoro action. The compatibility condition depends on the points $z_i$ and the local coordinates, see e.g.\ \cite{DiFr:1997} or \cite{Huang:1994,Frenkel:2004}. In the case of Virasoro minimal models the spaces $\mathrm{Bl}(\Sigma)$ are finite-dimensional.

A correlator $C$ on $\hat\Cb$ with insertions of bulk, defect, or junction fields $\psi_i \in V_i \otimes_\Cb \bar W_i$ at points $z_i$ with local coordinate $f_{z_i,\theta_i}$ is an element of $\mathrm{Bl}(\Sigma) \otimes_\Cb \mathrm{Bl}(\tilde\Sigma)$. Here $V_i$ and $W_i$ are representations of the Virasoro algebra as above, $\Sigma$ is as in \eqref{eq:Sigma-for-n-points} and 
\be
  \tilde\Sigma = \big( (z_1^*,-\theta_1,W_1),\dots,(z_n^*,-\theta_n,W_n) \big) \ .
\ee

The category of representations of the vertex operator algebra build on the vacuum representation $R_1$ is a so-called modular category \cite{Moore:1989vd,Turaev:1994}, which we denote by $\Cc$. The modular category $\Cc$ in turn defines a 3d\,TFT, namely a symmetric monoidal functor $\mathrm{tft}_\Cc$ from extended bordisms to complex vector spaces \cite{Turaev:1994}, see \cite[Sect.\,2]{tft1} for an introduction using the present notation. In particular, to an extended Riemann sphere $\Sigma$, the functor $\mathrm{tft}_\Cc$ assigns a vector space, which can be identified with $\mathrm{Bl}(\Sigma)$, 
\be
  \mathrm{tft}_\Cc(\Sigma) = \mathrm{Bl}(\Sigma) \ .
\ee
By monoidality, we also have $\mathrm{tft}_\Cc(\Sigma \sqcup \tilde\Sigma) = \mathrm{Bl}(\Sigma) \otimes_\Cb \mathrm{Bl}(\tilde\Sigma)$. Consider an extended bordism $M : \emptyset \rightarrow \Sigma \sqcup \tilde\Sigma$, i.e.\ a three-manifold whose boundary is $\Sigma \sqcup \tilde\Sigma$ and which has an embedded ribbon graph. To $M$, the 3d\,TFT assigns a linear map
\be
   \mathrm{tft}_\Cc(M) : \Cb \longrightarrow  \mathrm{Bl}(\Sigma) \otimes_\Cb \mathrm{Bl}(\tilde\Sigma) \ .
\ee
We can thus describe correlators of the CFT by constructing an appropriate extended bordism $M$ and taking the correlator to be the image of $1 \in \Cb$ under the map $\mathrm{tft}_\Cc(M)$. The construction of the extended bordism will be given below. More details on how to compute the expansion of $\mathrm{tft}_\Cc(M)$ in a given basis of conformal blocks can be found in \cite[Sect.\,5]{tft4}.

\subsection{Ribbon representation of bulk, boundary, and defect fields}

\begin{figure}[bt]

\begin{align*}
&  
a)~  \raisebox{-40pt}{
  \begin{picture}(60,58)
   \put(0,0){\scalebox{.75}{\includegraphics{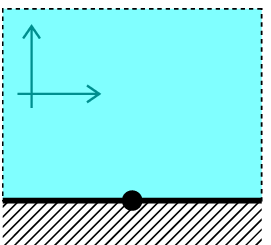}}}
   \put(0,0){
     \setlength{\unitlength}{.75pt}\put(-34,-18){
     \put (46, 76) {\scriptsize$ 2 $}
     \put (58, 67) {\scriptsize$ 1 $}
     \put (40, 36) {\scriptsize$ b $}
     \put (99, 36) {\scriptsize$ b $}
     \put (62, 38) {\scriptsize$ \psi^b(r) $}
     }\setlength{\unitlength}{1pt}}
  \end{picture}}
&\hspace{0.5em}&
b)~\raisebox{-40pt}{
  \begin{picture}(60,58)
   \put(0,0){\scalebox{.75}{\includegraphics{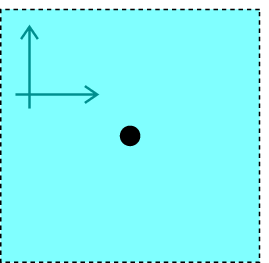}}}
   \put(0,0){
     \setlength{\unitlength}{.75pt}\put(-34,-15){
     \put (46, 78) {\scriptsize$ 2 $}
     \put (58, 69) {\scriptsize$ 1 $}
     \put (78, 45) {\scriptsize$ \Phi(z) $}
     }\setlength{\unitlength}{1pt}}
  \end{picture}}
\\[-3em]
&
\hspace{2.8em}\longmapsto
 \eta^{(b)}_\phi
   \raisebox{-35pt}{
  \begin{picture}(133,80)
   \put(0,0){\scalebox{.75}{\includegraphics{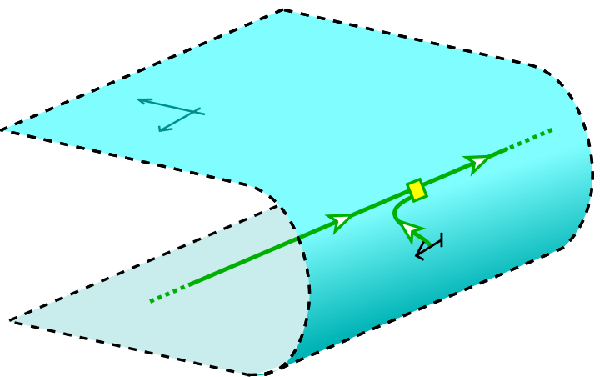}}}
   \put(0,0){
     \setlength{\unitlength}{.75pt}\put(-266,-58){
     \put (307,125) {\scriptsize$ 1 $}
     \put (299,135) {\scriptsize$ 2 $}
     \put (359,107) {\scriptsize$ b $}
     \put (399,124) {\scriptsize$ b $}
     \put (375,96) {\scriptsize$ 3 $}
     \put (396,96) {\scriptsize$ r $}
     }\setlength{\unitlength}{1pt}}
  \end{picture}}
&&
\hspace{1.5em}\mapsto
  e^{\pi i h_3} \dim(3) \, \eta_\Phi
  \raisebox{-45pt}{
  \begin{picture}(130,105)
   \put(0,0){\scalebox{.75}{\includegraphics{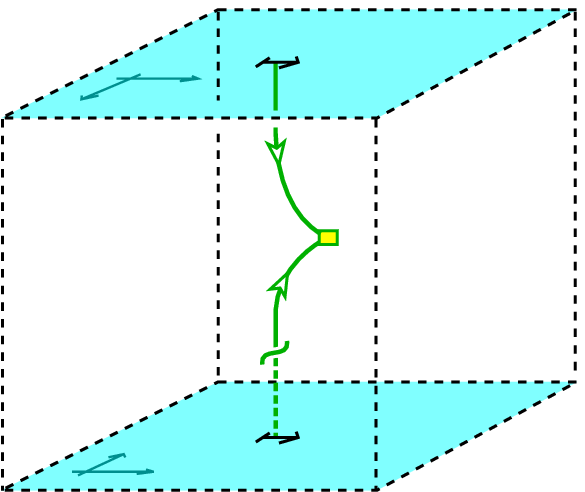}}}
   \put(0,0){
     \setlength{\unitlength}{.75pt}\put(-5,-3){
     \put (52,7) {\scriptsize$ 1 $}
     \put (42,14) {\scriptsize$ 2 $}
     \put (56,125) {\scriptsize$ 1 $}
     \put (36,113) {\scriptsize$ 2 $}
     \put (91,60) {\scriptsize$ 3 $}
     \put (90,95) {\scriptsize$ 3 $}
     \put (94,125) {\scriptsize$ z $}
     \put (94,16) {\scriptsize$ z^* $}
     }\setlength{\unitlength}{1pt}}
  \end{picture}}
\\[2em]
&
  c)~\raisebox{-40pt}{
  \begin{picture}(60,58)
   \put(0,0){\scalebox{.75}{\includegraphics{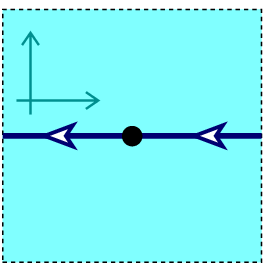}}}
   \put(0,0){
     \setlength{\unitlength}{.75pt}\put(-34,-15){
     \put (46, 76) {\scriptsize$ 2 $}
     \put (58, 67) {\scriptsize$ 1 $}
     \put (40, 42) {\scriptsize$ a $}
     \put (99, 42) {\scriptsize$ a $}
     \put (62, 38) {\scriptsize$ \phi^a(z) $}
     }\setlength{\unitlength}{1pt}}
  \end{picture}}
&&
  d)~\raisebox{-40pt}{
  \begin{picture}(60,58)
   \put(0,0){\scalebox{.75}{\includegraphics{pic16c.eps}}}
   \put(0,0){
     \setlength{\unitlength}{.75pt}\put(-34,-15){
     \put (46, 76) {\scriptsize$ 2 $}
     \put (58, 67) {\scriptsize$ 1 $}
     \put (40, 42) {\scriptsize$ a $}
     \put (99, 42) {\scriptsize$ a $}
     \put (62, 38) {\scriptsize$ \bar\phi^a(z) $}
     }\setlength{\unitlength}{1pt}}
  \end{picture}}
\\[-3em]  
&
\hspace{3.5em}\longmapsto
  \eta^{(a)}_\phi
  \raisebox{-45pt}{
  \begin{picture}(130,105)
   \put(0,0){\scalebox{.75}{\includegraphics{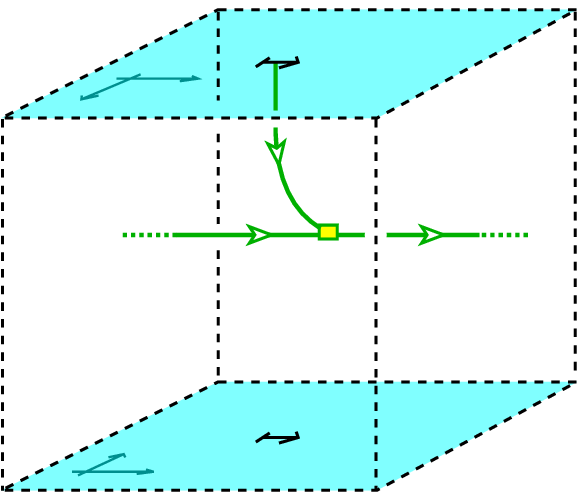}}}
   \put(0,0){
     \setlength{\unitlength}{.75pt}\put(-5,-3){
     \put (52,7) {\scriptsize$ 1 $}
     \put (42,14) {\scriptsize$ 2 $}
     \put (56,125) {\scriptsize$ 1 $}
     \put (36,113) {\scriptsize$ 2 $}
     \put (90,95) {\scriptsize$ 3 $}
     \put (74,67) {\scriptsize$ a $}
     \put (125,67) {\scriptsize$ a $}
     \put (94,125) {\scriptsize$ z $}
     \put (94,16) {\scriptsize$ z^* $}
     }\setlength{\unitlength}{1pt}}
  \end{picture}}
&&
\hspace{3em}\longmapsto
  e^{\pi i h_3} \, \eta^{(a)}_\phi
  \raisebox{-45pt}{
  \begin{picture}(130,105)
   \put(0,0){\scalebox{.75}{\includegraphics{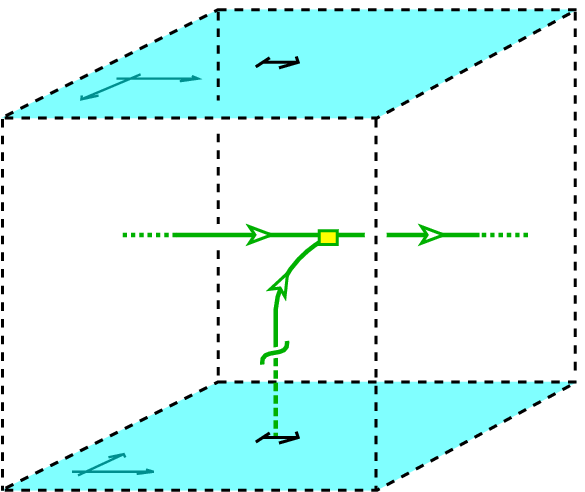}}}
   \put(0,0){
     \setlength{\unitlength}{.75pt}\put(-5,-3){
     \put (52,7) {\scriptsize$ 1 $}
     \put (42,14) {\scriptsize$ 2 $}
     \put (56,125) {\scriptsize$ 1 $}
     \put (36,113) {\scriptsize$ 2 $}
     \put (89,56) {\scriptsize$ 3 $}
     \put (74,67) {\scriptsize$ a $}
     \put (125,67) {\scriptsize$ a $}
     \put (94,125) {\scriptsize$ z $}
     \put (94,16) {\scriptsize$ z^* $}
     }\setlength{\unitlength}{1pt}}
  \end{picture}}
\end{align*}

\caption{Ribbon graph representations: 
(a) boundary field $\psi^b$;
(b) bulk field $\Phi$; 
(c) holomorphic defect field $\phi^a$;
(d) anti-holomorphic defect field $\bar\phi^a$.
Recall that $3 \equiv (1,3)$.}
\label{fig:field-ins}
\end{figure}

A ribbon graph inside a bordism $M$ consists of ribbons which can end at the marked points on the boundary of $M$, or at coupons in the interior of $M$. A ribbon has an orientation of its surface and a direction, and it is labelled by an object of $\Cc$, i.e.\ by a representation of the Virasoro vertex operator algebra $R_1$. The coupons are labelled by morphisms in $\Cc$, i.e.\ by intertwiners between fusion products of representations.

When drawing pictures of ribbon graphs, we draw solid lines to indicate that the ribbon carries the orientation of the paper plane, and dashed lines to indicate the opposite orientation. The direction of the ribbon is indicated by an arrow, see \cite[Sect.\,2]{tft1} for more on these conventions.

We will only need coupons with two incoming and one outgoing ribbons, and vice versa. For incoming ribbons labelled $R_i$ and $R_j$ and the outgoing ribbon labelled $R_k$, the coupon is labelled by an element in $\Hom_\Cc(R_i \otimes R_j , R_k)$, i.e.\ an intertwiner from $R_i \times R_j$ to $R_k$. We pick a basis element $\lambda_{ij}^k \in \Hom_\Cc(R_i \otimes R_j , R_k)$ and use it to label such coupons. We also choose the dual basis $\bar \lambda^{ij}_k \in  \Hom_\Cc( R_k, R_i \otimes R_j)$, such that $\lambda_{ij}^k \circ \bar \lambda^{ij}_k = \id_{R_k}$. These two coupons will be drawn as little squares,
\be
  \raisebox{-35pt}{
  \begin{picture}(30,70)
   \put(0,8){\scalebox{.75}{\includegraphics{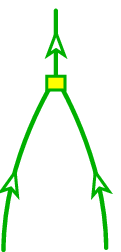}}}
   \put(0,8){
     \setlength{\unitlength}{.90pt}\put(-16,-16){
     \put( 11,  22) {\scriptsize$ i $}
     \put( 43,  22) {\scriptsize$ j $}
     \put( 20, 65) {\scriptsize$ k $}
     \put( 34, 55) {\scriptsize$ \lambda_{ij}^k $}
     }\setlength{\unitlength}{1pt}}
  \end{picture}}
\qquad , \qquad
  \raisebox{-35pt}{
  \begin{picture}(30,70)
   \put(0,8){\scalebox{.75}{\includegraphics{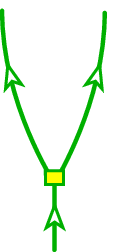}}}
   \put(0,8){
     \setlength{\unitlength}{.90pt}\put(-16,-16){
     \put( 11,  65) {\scriptsize$ i $}
     \put( 43,  65) {\scriptsize$ j $}
     \put( 20, 22) {\scriptsize$ k $}
     \put( 34, 30) {\scriptsize$\bar \lambda^{ij}_k $}
     }\setlength{\unitlength}{1pt}}
  \end{picture}}
\quad . 
\ee
The linear map $\mathrm{tft}_\Cc(M)$ associated to a bordism $M$ does not change if we execute any of the modifications below on a part of the embedded ribbon graph:
\begin{align}
  \raisebox{-42pt}{
  \begin{picture}(70,85)
   \put(0,8){\scalebox{.75}{\includegraphics{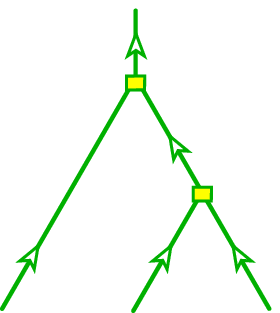}}}
   \put(0,8){
     \setlength{\unitlength}{.90pt}\put(-16,-16){
     \put( 16,  28) {\scriptsize$ i $}
     \put( 47,  28) {\scriptsize$ j $}
     \put( 78,  28) {\scriptsize$ k $}
     \put( 60, 62) {\scriptsize$ p $}
     \put( 42, 79) {\scriptsize$ l $}
     }\setlength{\unitlength}{1pt}}
  \end{picture}}
  \hspace{-1em}
  &= ~ \sum_{q\in\Ic} ~
  \Fs^{(ijk)l}_{pq} 
  \hspace{-.5em}
  \raisebox{-42pt}{
  \begin{picture}(70,85)
   \put(0,8){\scalebox{.75}{\includegraphics{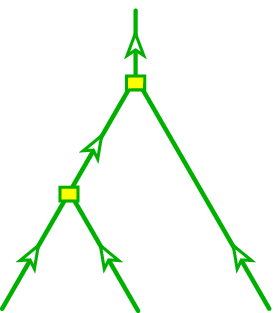}}}
   \put(0,8){
     \setlength{\unitlength}{.90pt}\put(-16,-16){
     \put( 16,  28) {\scriptsize$ i $}
     \put( 47,  28) {\scriptsize$ j $}
     \put( 78,  28) {\scriptsize$ k $}
     \put( 33, 62) {\scriptsize$ q $}
     \put( 42, 79) {\scriptsize$ l $}
     }\setlength{\unitlength}{1pt}}
  \end{picture}}  
&
  \raisebox{-42pt}{
  \begin{picture}(70,85)
   \put(0,8){\scalebox{.75}{\includegraphics{pic18b.eps}}}
   \put(0,8){
     \setlength{\unitlength}{.90pt}\put(-16,-16){
     \put( 16,  28) {\scriptsize$ i $}
     \put( 47,  28) {\scriptsize$ j $}
     \put( 78,  28) {\scriptsize$ k $}
     \put( 33, 62) {\scriptsize$ q $}
     \put( 42, 79) {\scriptsize$ l $}
     }\setlength{\unitlength}{1pt}}
  \end{picture}}  
  \hspace{-1em}
  &= ~ \sum_{p\in\Ic} ~
  \Gs^{(ijk)l}_{qp} 
  \hspace{-.5em}
  \raisebox{-42pt}{
  \begin{picture}(70,85)
   \put(0,8){\scalebox{.75}{\includegraphics{pic18a.eps}}}
   \put(0,8){
     \setlength{\unitlength}{.90pt}\put(-16,-16){
     \put( 16,  28) {\scriptsize$ i $}
     \put( 47,  28) {\scriptsize$ j $}
     \put( 78,  28) {\scriptsize$ k $}
     \put( 60, 62) {\scriptsize$ p $}
     \put( 42, 79) {\scriptsize$ l $}
     }\setlength{\unitlength}{1pt}}
  \end{picture}}
\nonumber\\
  \raisebox{-35pt}{
  \begin{picture}(30,70)
   \put(0,8){\scalebox{.75}{\includegraphics{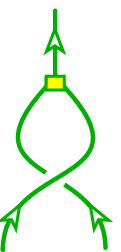}}}
   \put(0,8){
     \setlength{\unitlength}{.90pt}\put(-16,-16){
     \put( 11,  22) {\scriptsize$ i $}
     \put( 43,  22) {\scriptsize$ j $}
     \put( 20, 65) {\scriptsize$ k $}
     }\setlength{\unitlength}{1pt}}
  \end{picture}}
  &=~ \Rs^{(ij)k}
  \raisebox{-35pt}{
  \begin{picture}(30,70)
   \put(0,8){\scalebox{.75}{\includegraphics{pic18d.eps}}}
   \put(0,8){
     \setlength{\unitlength}{.90pt}\put(-16,-16){
     \put( 11,  22) {\scriptsize$ i $}
     \put( 43,  22) {\scriptsize$ j $}
     \put( 20, 65) {\scriptsize$ k $}
     }\setlength{\unitlength}{1pt}}
  \end{picture}}
&
  \raisebox{-35pt}{
  \begin{picture}(30,70)
   \put(0,8){\scalebox{.75}{\includegraphics{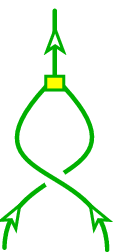}}}
   \put(0,8){
     \setlength{\unitlength}{.90pt}\put(-16,-16){
     \put( 11,  22) {\scriptsize$ i $}
     \put( 43,  22) {\scriptsize$ j $}
     \put( 20, 65) {\scriptsize$ k $}
     }\setlength{\unitlength}{1pt}}
  \end{picture}}
  &=~ \frac{1}{\Rs^{(ji)k}}
  \raisebox{-35pt}{
  \begin{picture}(30,70)
   \put(0,8){\scalebox{.75}{\includegraphics{pic18d.eps}}}
   \put(0,8){
     \setlength{\unitlength}{.90pt}\put(-16,-16){
     \put( 11,  22) {\scriptsize$ i $}
     \put( 43,  22) {\scriptsize$ j $}
     \put( 20, 65) {\scriptsize$ k $}
     }\setlength{\unitlength}{1pt}}
  \end{picture}}
\label{eq:local-ribbon-moves}\\
  \raisebox{-35pt}{
  \begin{picture}(30,80)
   \put(0,8){\scalebox{.75}{\includegraphics{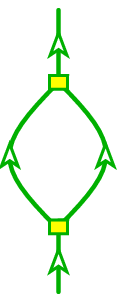}}}
   \put(0,8){
     \setlength{\unitlength}{.90pt}\put(-16,-16){
     \put( 11,  48) {\scriptsize$ k $}
     \put( 45,  48) {\scriptsize$ l $}
     \put( 24, 75) {\scriptsize$ j $}
     \put( 24, 20) {\scriptsize$ i $}
     }\setlength{\unitlength}{1pt}}
  \end{picture}} ~
&= ~ \delta_{i,j} \, N_{kl}^{~i} ~~
  \raisebox{-35pt}{
  \begin{picture}(5,80)
   \put(0,8){\scalebox{.75}{\includegraphics{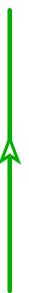}}}
   \put(0,8){
     \setlength{\unitlength}{.90pt}\put(-16,-16){
     \put( 12,  48) {\scriptsize$ i $}
     }\setlength{\unitlength}{1pt}}
  \end{picture}}
&
  \raisebox{-17pt}{
  \begin{picture}(30,30)
   \put(0,8){\scalebox{.75}{\includegraphics{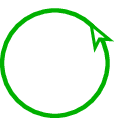}}}
   \put(0,8){
     \setlength{\unitlength}{.90pt}\put(-16,-16){
     \put( 43,  22) {\scriptsize$ i $}
     }\setlength{\unitlength}{1pt}}
  \end{picture}}
&= ~ \dim(i)
\nonumber
\end{align}
It is understood that the sums are taken after applying $\mathrm{tft}_\Cc(-)$.

If the surface is the Riemann sphere $\hat\Cb$, the relevant three-manifold is $M_{\hat\Cb} = \hat\Cb \times [-1,1]$. If the surface is the upper half plane with infinity, $\hat\Hb = \{ z \,|\, \text{im}(z) {\ge} 0 \} \cup \{ \infty \}$, the three manifold is given by $M_{\hat\Hb} = \hat\Hb \times [-1,1] / \sim$, where $(x,t) \sim (x,-t)$ for $x \in \Rb \cup \{ \infty \}$. The boundary of $M_{\hat\Cb}$ is isomorphic to $\hat\Cb \sqcup \hat\Cb$, and the boundary of $M_{\hat\Hb}$ is isomorphic to $\hat\Cb$.

With this notation in place, the ribbon graphs representing the various field insertions are given in Figure~\ref{fig:field-ins}, following \cite[Sect.\,3]{tft4}. Shown in these pictures is a fraction of the surface $X$ containing a field insertion, and a fraction of the three-manifold $M$ with the corresponding ribbon graph. The normalisation constants are included for convenience. The holomorphic coordinate is associated to the `top' boundary of the fraction of $M$ drawn, and the orientation between defect lines and the ribbons that represent them gets reversed (this convention is chosen to match \cite{tft4}). Because the boundary of $M$ is oriented by the inward pointing normal, the surface $X$ gets mapped to this boundary by an orientation reversing map. This is the reason for the reflection relating the defect graph inside $X$ and the ribbon graph inside $M$ noticeable in the examples in the next section.

\subsection{Calculation of structure constants}\label{sec:TFT-struct}

\subsubsection*{OPE of bulk fields}

The OPE coefficients $C_{\Phi\Phi}^{~\one}$ and $C_{\Phi\Phi}^{~\Phi}$ in \eqref{eq:bulk-structure-constants} and \eqref{eq:structure-const-via-F} are found from the 3d\,TFT representation as follows.
\bea
  \Phi(z)\Phi(w) 
\\[-1em]\displaystyle
   \longmapsto
  \big( e^{\pi i h_3} \dim(3) \,\eta_\Phi \big)^2 ~
  \raisebox{-35pt}{
  \begin{picture}(159,105)
   \put(0,0){\scalebox{.75}{\includegraphics{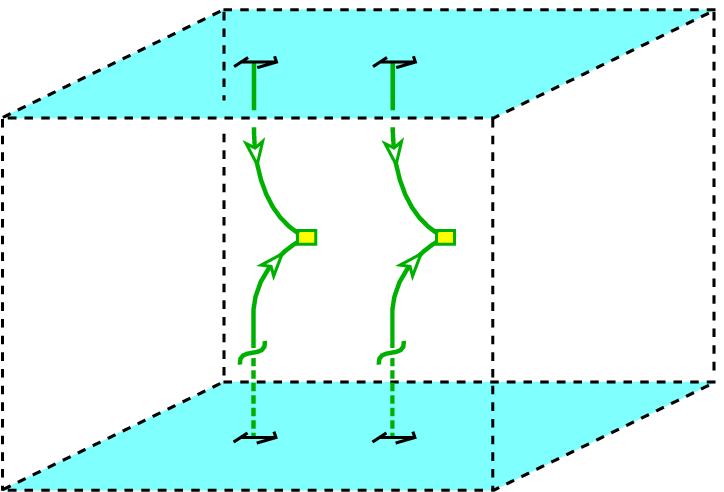}}}
   \put(0,0){
     \setlength{\unitlength}{.75pt}\put(-5,-7){
     \put (127,128) {\scriptsize$ z $}
     \put (127,20) {\scriptsize$ z^* $}
     \put (88,128) {\scriptsize$ w $}
     \put (88,20) {\scriptsize$ w^* $}
     \put (83,99) {\scriptsize$ 3 $}
     \put (83,59) {\scriptsize$ 3 $}
     \put (123,99) {\scriptsize$ 3 $}
     \put (123,59) {\scriptsize$ 3 $}
     }\setlength{\unitlength}{1pt}}
  \end{picture}}
\\[-1.5em]\displaystyle
\hspace{9em}
=  \big( e^{\pi i h_3} \dim(3) \eta_\Phi \big)^2 \sum_k D_{\Phi\Phi}^k 
  \raisebox{-35pt}{
  \begin{picture}(159,105)
   \put(0,0){\scalebox{.75}{\includegraphics{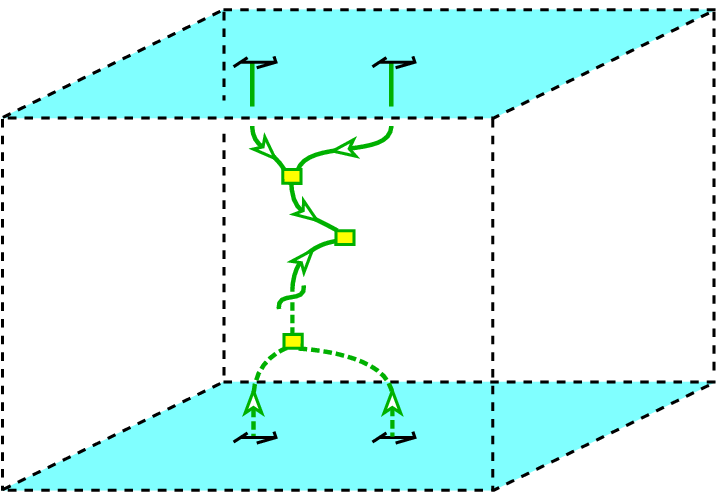}}}
   \put(0,0){
     \setlength{\unitlength}{.75pt}\put(-5,-7){
     \put (127,128) {\scriptsize$ z $}
     \put (127,20) {\scriptsize$ z^* $}
     \put (88,128) {\scriptsize$ w $}
     \put (88,20) {\scriptsize$ w^* $}
     \put (70,120) {\scriptsize$ 3 $}
     \put (110,120) {\scriptsize$ 3 $}
     \put (69,28) {\scriptsize$ 3 $}
     \put (109,28) {\scriptsize$ 3 $}
     \put (83,83) {\scriptsize$ k $}
     \put (95,63) {\scriptsize$ k $}
     }\setlength{\unitlength}{1pt}}
  \end{picture}}
\\[4em]\displaystyle
\longmapsto
\big( e^{\pi i h_3} \dim(3) \, \eta_\Phi \big)^2 D_{\Phi\Phi}^1 |z-w|^{-4h_3} \one
+ e^{\pi i h_3} \dim(3) \eta_\Phi  D_{\Phi\Phi}^3 |z-w|^{-2h_3} \Phi(w) + \dots
\eear\labl{eq:bulk-ope-aux1}
The constants $D_{\Phi\Phi}^k$ are determined by the identity
\be
  \raisebox{-25pt}{
  \begin{picture}(75,65)
   \put(0,0){\scalebox{.75}{\includegraphics{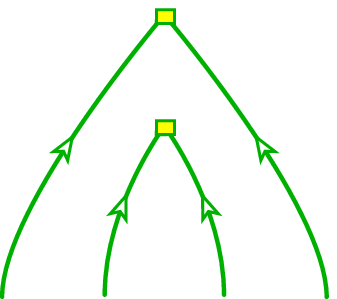}}}
   \put(0,0){
     \setlength{\unitlength}{.75pt}\put(4,-16){
     \put (3,28) {\scriptsize$ 3 $}
     \put (31,28) {\scriptsize$ 3 $}
     \put (50,28) {\scriptsize$ 3 $}
     \put (78,28) {\scriptsize$ 3 $}
     }\setlength{\unitlength}{1pt}}
  \end{picture}}
  ~=~\sum_{k \in \Ic} D_{\Phi\Phi}^k~
  \raisebox{-25pt}{
  \begin{picture}(75,65)
   \put(0,0){\scalebox{.75}{\includegraphics{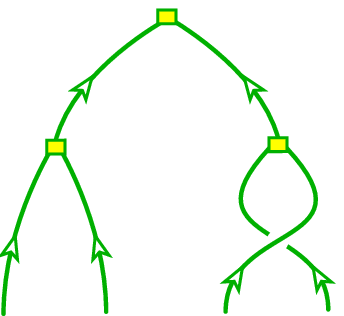}}}
   \put(0,0){
     \setlength{\unitlength}{.75pt}\put(4,-16){
     \put (3,28) {\scriptsize$ 3 $}
     \put (30,28) {\scriptsize$ 3 $}
     \put (54,28) {\scriptsize$ 3 $}
     \put (92,28) {\scriptsize$ 3 $}
     \put (26,78) {\scriptsize$ k $}
     \put (58,78) {\scriptsize$ k $}
     }\setlength{\unitlength}{1pt}}
  \end{picture}}
  \quad,
\ee
see \cite[Sect.\,2.2]{tft4} and in particular \cite[Eqn.\,(2.26)]{tft4}. Composing both sides from below with the morphism dual to the right hand side, applying the local moves \eqref{eq:local-ribbon-moves} and using \eqref{eq:mm-Rs-val} and \eqref{eq:F1p-vs-Fp1} gives
\be
  D_{\Phi\Phi}^k = 
  \raisebox{-50pt}{
  \begin{picture}(75,100)
   \put(0,0){\scalebox{.75}{\includegraphics{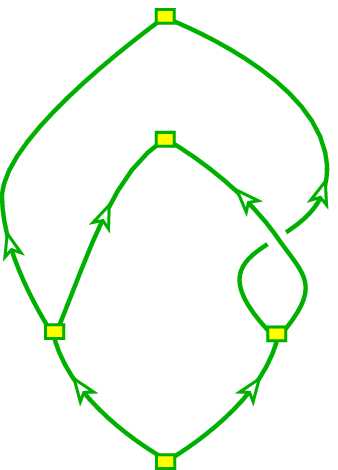}}}
   \put(0,0){
     \setlength{\unitlength}{.75pt}\put(4,-16){
     \put (29,33) {\scriptsize$ k $}
     \put (53,33) {\scriptsize$ k $}
     \put (1,96) {\scriptsize$ 3 $}
     \put (32,90) {\scriptsize$ 3 $}
     \put (56,90) {\scriptsize$ 3 $}
     \put (80,96) {\scriptsize$ 3 $}
     }\setlength{\unitlength}{1pt}}
  \end{picture}}
  =  \frac{\Fs^{(333)3}_{1k}}{\Rs^{(33)k}}
  = e^{\pi i (- 2 h_3 + h_k)} \frac{\dim(k)}{\dim(3)^2} \frac{1}{\Fs^{(333)3}_{k1}} \ .
\ee
Substituting this into \eqref{eq:bulk-ope-aux1}, we recover \eqref{eq:structure-const-via-F}.
  
\subsubsection*{OPE of defect fields and boundary fields}

The OPE of the holomorphic defect fields $\phi^s$ was already computed in \cite[App.\,A.2]{Runkel:2007wd}, and one finds the constants $C^{(s)\one}_{\phi\phi}$ and $C^{(s)\phi}_{\phi\phi}$ in \eqref{eq:structure-const-via-F}. It remains to check that the same constants appear in the OPE of the anti-holomorphic defect fields $\bar\phi^s$ and of the boundary fields $\psi^s$. We give the first two steps in each computation.
\bea
  \bar\phi^s(x) \bar\phi^s(y) 
\\[-1em]\displaystyle
\quad  \longmapsto 
  e^{2 \pi i h_3} \big(\eta^{(s)}_\phi\big)^2 
  \raisebox{-35pt}{
  \begin{picture}(159,105)
   \put(0,0){\scalebox{.75}{\includegraphics{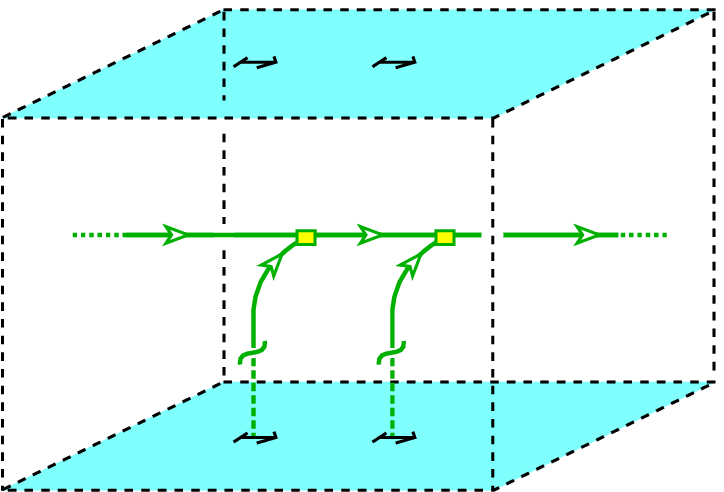}}}
   \put(0,0){
     \setlength{\unitlength}{.75pt}\put(-5,-7){
     \put (127,128) {\scriptsize$ x $}
     \put (127,20) {\scriptsize$ x $}
     \put (87,128) {\scriptsize$ y $}
     \put (87,20) {\scriptsize$ y $}
     \put (53,86) {\scriptsize$ s $}
     \put (109,86) {\scriptsize$ s $}
     \put (170,86) {\scriptsize$ s $}
     \put (81,59) {\scriptsize$ 3 $}
     \put (122,59) {\scriptsize$ 3 $}
     }\setlength{\unitlength}{1pt}}
  \end{picture}}
\\[-2em]\displaystyle
\hspace{11em}
  = e^{2 \pi i h_3} \big(\eta^{(s)}_\phi\big)^2 \sum_k \frac{\Fs^{(33s)s}_{sk}}{\Rs^{(33)k}} 
  \raisebox{-35pt}{
  \begin{picture}(159,105)
   \put(0,0){\scalebox{.75}{\includegraphics{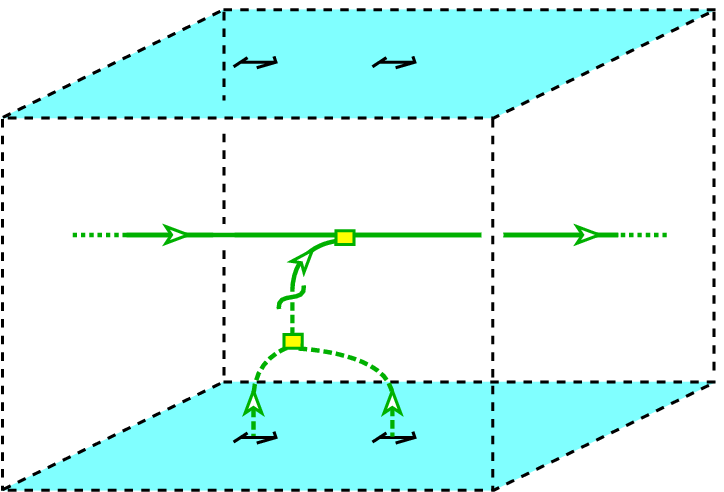}}}
   \put(0,0){
     \setlength{\unitlength}{.75pt}\put(-5,-7){
     \put (127,128) {\scriptsize$ x $}
     \put (127,20) {\scriptsize$ x $}
     \put (87,128) {\scriptsize$ y $}
     \put (87,20) {\scriptsize$ y $}
     \put (53,86) {\scriptsize$ s $}
     \put (170,86) {\scriptsize$ s $}
     \put (69,28) {\scriptsize$ 3 $}
     \put (109,28) {\scriptsize$ 3 $}
     \put (95,63) {\scriptsize$ k $}
     }\setlength{\unitlength}{1pt}}
  \end{picture}}
  \ , 
\eear\ee
and
\bea
  \psi^s(x) \psi^s(y) 
\\[-1em]\displaystyle
 \quad \longmapsto 
  \big(\eta^{(s)}_\phi\big)^2 
  \raisebox{-35pt}{
  \begin{picture}(135,80)
   \put(0,0){\scalebox{.75}{\includegraphics{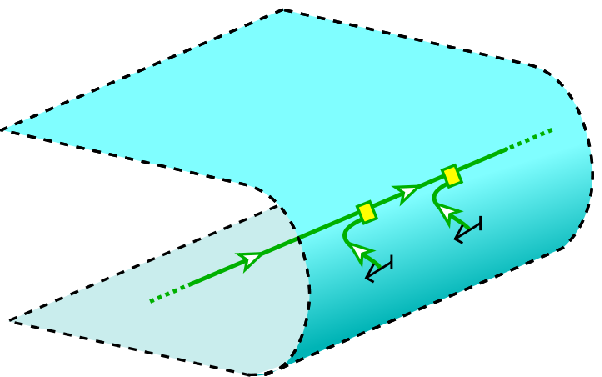}}}
   \put(0,0){
     \setlength{\unitlength}{.75pt}\put(-265,-59){
     \put (334,97) {\scriptsize$ s $}
     \put (376,116) {\scriptsize$ s $}
     \put (403,125) {\scriptsize$ s $}
     \put (372,81) {\scriptsize$ x $}
     \put (398,92) {\scriptsize$ y $}
     \put (360,90) {\scriptsize$ 3 $}
     \put (387,100) {\scriptsize$ 3 $}
     }\setlength{\unitlength}{1pt}}
  \end{picture}}
  = \big(\eta^{(s)}_\phi\big)^2 \sum_k \Fs^{(33s)s}_{sk} 
  \raisebox{-35pt}{
  \begin{picture}(135,80)
   \put(0,0){\scalebox{.75}{\includegraphics{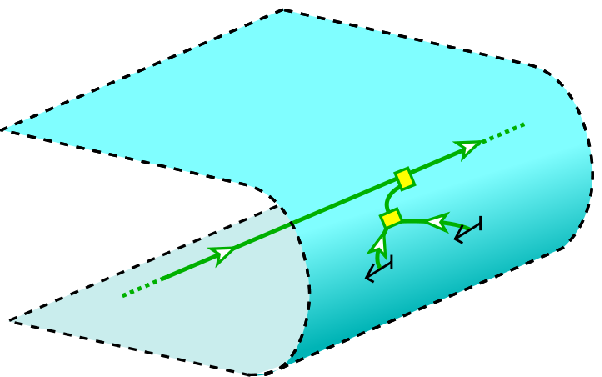}}}
   \put(0,0){
     \setlength{\unitlength}{.75pt}\put(-265,-59){
     \put (334,90) {\scriptsize$ s $}
     \put (403,120) {\scriptsize$ s $}
     \put (372,81) {\scriptsize$ x $}
     \put (398,92) {\scriptsize$ y $}
     \put (370,108) {\scriptsize$ k $}
     \put (365,93) {\scriptsize$ 3 $}
     \put (390,108) {\scriptsize$ 3 $}
     }\setlength{\unitlength}{1pt}}
  \end{picture}}
  \ .
\eear\ee

\subsubsection*{Bubble collapse with defect fields}

The coefficients in \eqref{eq:bubble-collapse} involving holomorphic defect fields have been computed in \cite[App.\,A.2]{Runkel:2007wd}. For the identities involving anti-holomorphic defect fields, we only give the corresponding bordism, the constants are obtained using the moves \eqref{eq:local-ribbon-moves}.
\bea
\\
  \jun^{d \rightarrow b\star e} ~ (\bar\phi^{b\rightarrow a} \times \one^e)(x) ~ \bar\jun^{a\star e \rightarrow c} 
\\[-2em]
  \hspace{3em} =~
  \raisebox{-25pt}{
  \begin{picture}(100,58)
   \put(0,0){\scalebox{.75}{\includegraphics{pic05a.eps}}}
   \put(0,0){
     \setlength{\unitlength}{.75pt}\put(-34,-15){
     \put( 45, 58) {\scriptsize$ c $}
     \put(142, 58) {\scriptsize$ d $}
     \put( 79, 58) {\scriptsize$ a $}
     \put(114, 58) {\scriptsize$ b $}
     \put(117, 76) {\scriptsize$ e $}
     \put( 88, 36) {\scriptsize$ \bar\phi^{a \leftarrow b} $}
     }\setlength{\unitlength}{1pt}}
  \end{picture}}
  \quad\longmapsto\quad  
  e^{\pi i h_f} \, \eta^{ab} ~
  \raisebox{-35pt}{
  \begin{picture}(159,105)
   \put(0,0){\scalebox{.75}{\includegraphics{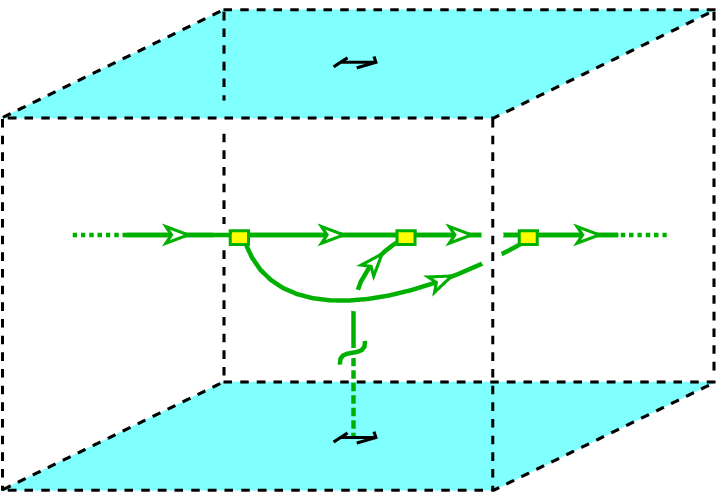}}}
   \put(0,0){
     \setlength{\unitlength}{.75pt}\put(-5,-7){
     \put (53,87) {\scriptsize$ c $}
     \put (98,87) {\scriptsize$ a $}
     \put (132,87) {\scriptsize$ b $}
     \put (170,87) {\scriptsize$ d $}
     \put (130,59) {\scriptsize$ e $}
     \put (98,28) {\scriptsize$ f $}
     }\setlength{\unitlength}{1pt}}
  \end{picture}}
  \\[5em]\displaystyle
  \jun^{d \rightarrow e\star b} ~ (\one^e \times \bar\phi^{b\rightarrow a})(x) ~ \bar\jun^{e\star a \rightarrow c} 
  \\[-2em]
  \hspace{3em} =~
  \raisebox{-25pt}{
  \begin{picture}(102,58)
   \put(0,0){\scalebox{.75}{\includegraphics{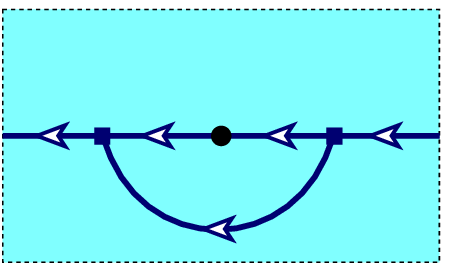}}}
   \put(0,0){
     \setlength{\unitlength}{.75pt}\put(-34,-15){
     \put( 45, 58) {\scriptsize$ c $}
     \put(142, 58) {\scriptsize$ d $}
     \put( 79, 58) {\scriptsize$ a $}
     \put(114, 58) {\scriptsize$ b $}
     \put(117, 23) {\scriptsize$ e $}
     \put( 88, 37) {\scriptsize$ \bar\phi^{a \leftarrow b} $}
     }\setlength{\unitlength}{1pt}}
  \end{picture}}
  \quad\longmapsto\quad  
  e^{\pi i h_f} \, \eta^{ab} ~
  \raisebox{-35pt}{
  \begin{picture}(159,105)
   \put(0,0){\scalebox{.75}{\includegraphics{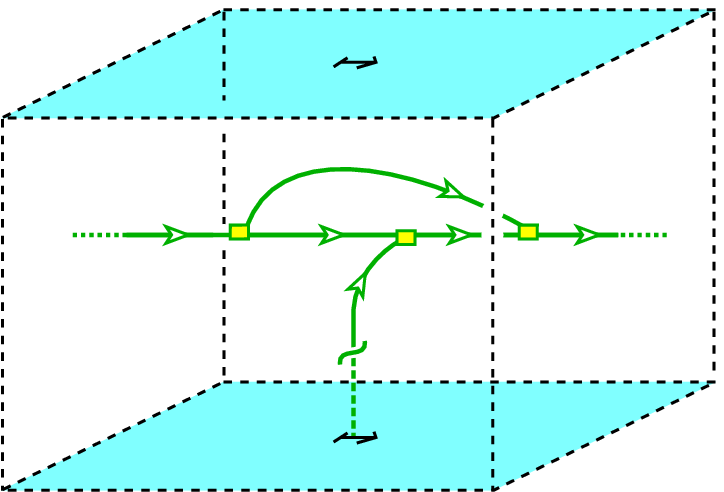}}}
   \put(0,0){
     \setlength{\unitlength}{.75pt}\put(-5,-7){
     \put (53,71) {\scriptsize$ c $}
     \put (98,71) {\scriptsize$ a $}
     \put (132,69) {\scriptsize$ b $}
     \put (170,69) {\scriptsize$ d $}
     \put (130,100) {\scriptsize$ e $}
     \put (98,28) {\scriptsize$ f $}
     }\setlength{\unitlength}{1pt}}
  \end{picture}}
\eear\ee

\subsubsection*{Defect circle with defect field}

We compute the coefficient $C_s$ in \eqref{eq:phiphi-circle-coefficient}. The ribbon graph representation of the left hand sides in \eqref{eq:phibarphi-to-Phi} is given by
\bea
  \raisebox{-25pt}{
  \begin{picture}(60,58)
   \put(0,0){\scalebox{.75}{\includegraphics{pic08a.eps}}}
   \put(0,0){
     \setlength{\unitlength}{.75pt}\put(-34,-15){
     \put (58, 25) {\scriptsize$ (1{,}s) $}
     \put (59, 59) {\scriptsize$ \phi\bar\phi^s(x) $}
     }\setlength{\unitlength}{1pt}}
  \end{picture}}
  \longmapsto e^{\pi i h_3} \big(\eta^{(s)}_\phi\big)^2 
  \raisebox{-35pt}{
  \begin{picture}(130,105)
   \put(0,0){\scalebox{.75}{\includegraphics{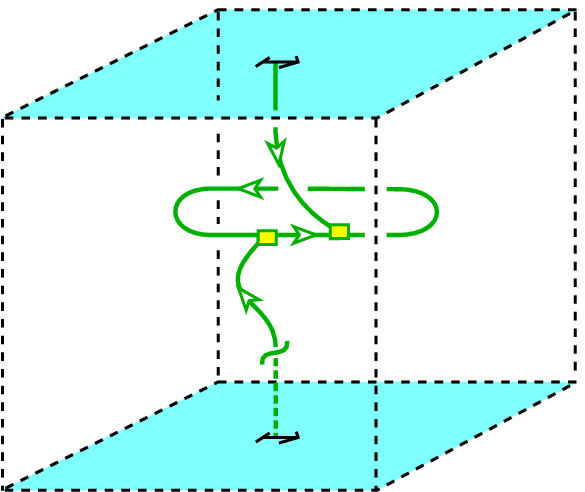}}}
   \put(0,0){
     \setlength{\unitlength}{.75pt}\put(-5,-3){
     \put (91,69) {\scriptsize$ s $}
     \put (51,89) {\scriptsize$ s $}
     \put (82,56) {\scriptsize$ 3 $}
     \put (89,101) {\scriptsize$ 3 $}
     }\setlength{\unitlength}{1pt}}
  \end{picture}}
\quad , 
\\[4em]\displaystyle
    \raisebox{-25pt}{
  \begin{picture}(60,58)
   \put(0,0){\scalebox{.75}{\includegraphics{pic08c.eps}}}
   \put(0,0){
     \setlength{\unitlength}{.75pt}\put(-34,-15){
     \put (58, 73) {\scriptsize$ (1{,}s) $}
     \put (59, 38) {\scriptsize$ \phi\bar\phi^s(x) $}
     }\setlength{\unitlength}{1pt}}
  \end{picture}} 
  \longmapsto e^{\pi i h_3} \big(\eta^{(s)}_\phi\big)^2 
  \raisebox{-35pt}{
  \begin{picture}(130,105)
   \put(0,0){\scalebox{.75}{\includegraphics{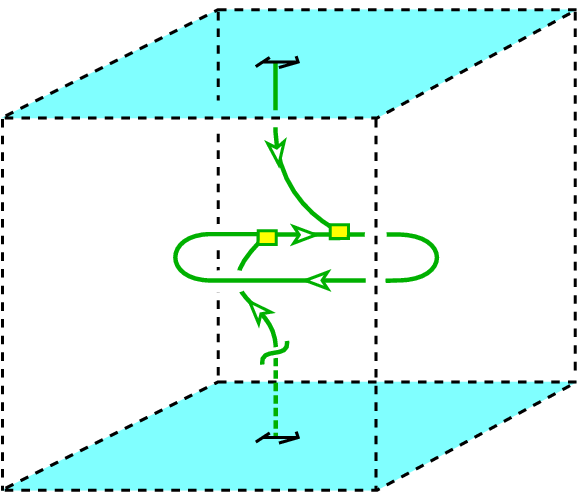}}}
   \put(0,0){
     \setlength{\unitlength}{.75pt}\put(-5,-3){
     \put (91,69) {\scriptsize$ s $}
     \put (95,55) {\scriptsize$ s $}
     \put (73,45) {\scriptsize$ 3 $}
     \put (89,101) {\scriptsize$ 3 $}
     }\setlength{\unitlength}{1pt}}
  \end{picture}}
  \quad . 
\eear\ee
Next one checks that
\be
  \raisebox{-42pt}{
  \begin{picture}(60,85)
   \put(0,8){\scalebox{.75}{\includegraphics{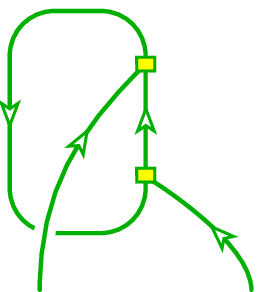}}}
   \put(0,8){
     \setlength{\unitlength}{.75pt}\put(-12,-17){
     \put( 29,  65) {\scriptsize$ 3 $}
     \put( 58,  66) {\scriptsize$ s$}
     \put( 77,  37) {\scriptsize$ 3 $}
     \put( 31,  90) {\scriptsize$ s$}
     }\setlength{\unitlength}{1pt}}
  \end{picture}}
  = ~B_s  ~
  \raisebox{-35pt}{
  \begin{picture}(30,70)
   \put(0,8){\scalebox{.75}{\includegraphics{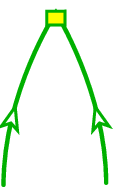}}}
   \put(0,8){
     \setlength{\unitlength}{.90pt}\put(-16,-16){
     \put( 11,  22) {\scriptsize$ 3 $}
     \put( 43,  22) {\scriptsize$ 3$}
     }\setlength{\unitlength}{1pt}}
  \end{picture}}
  \quad , \quad
  \raisebox{-42pt}{
  \begin{picture}(60,85)
   \put(0,8){\scalebox{.75}{\includegraphics{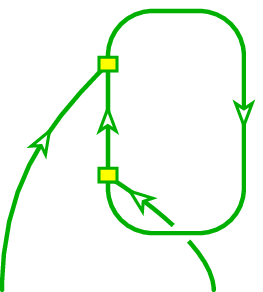}}}
   \put(0,8){
     \setlength{\unitlength}{.75pt}\put(-22,-17){
     \put( 28,  64) {\scriptsize$ 3 $}
     \put( 57,  66) {\scriptsize$ s$}
     \put( 65,  47) {\scriptsize$ 3 $}
     \put( 73,  90) {\scriptsize$ s$}
     }\setlength{\unitlength}{1pt}}
  \end{picture}}
  = ~(B_s)^* 
  \raisebox{-35pt}{
  \begin{picture}(30,70)
   \put(0,8){\scalebox{.75}{\includegraphics{pic30c.eps}}}
   \put(0,8){
     \setlength{\unitlength}{.90pt}\put(-16,-16){
     \put( 11,  22) {\scriptsize$ 3 $}
     \put( 43,  22) {\scriptsize$ 3$}
     }\setlength{\unitlength}{1pt}}
  \end{picture}}
\quad ,
\ee
with $B_s = \Rs^{(3s)s} \, \Fs^{(33s)s}_{s1}\dim(s)$ and $s \equiv (1,s)$. Together with the normalisation coefficient in Figure~\ref{fig:field-ins}b, this shows that
\be
  C_s = \frac{e^{\pi i h_3} \big(\eta^{(s)}_\phi\big)^2}{e^{\pi i h_3} \dim(3) \eta_\Phi} B_s \ ,
\ee
which is just \eqref{eq:phiphi-circle-coefficient}. 

\subsubsection*{Commutator with bulk field}

The left hand side of \eqref{eq:Phi-D2-comm} has ribbon representation
\be
  L = e^{\pi i h_3} \dim(3) \, \eta_\Phi \,
  \Bigg(   
  \raisebox{-35pt}{
  \begin{picture}(130,105)
   \put(0,0){\scalebox{.75}{\includegraphics{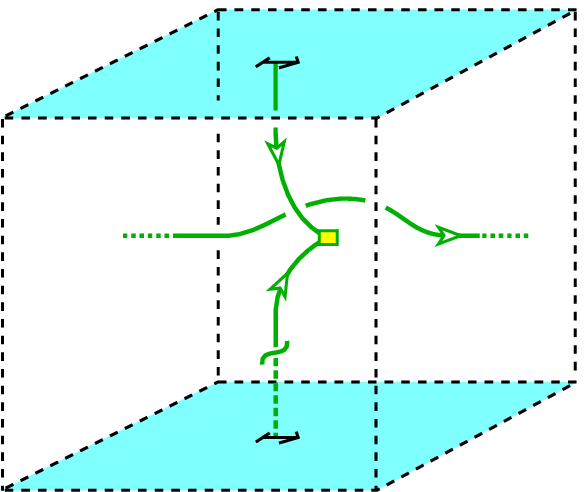}}}
   \put(0,0){
     \setlength{\unitlength}{.75pt}\put(-5,-3){
     \put (132,82) {\scriptsize$ 2 $}
     \put (75,99) {\scriptsize$ 3 $}
     \put (76,60) {\scriptsize$ 3 $}
     }\setlength{\unitlength}{1pt}}
  \end{picture}}
  ~-~
  \raisebox{-35pt}{
  \begin{picture}(130,105)
   \put(0,0){\scalebox{.75}{\includegraphics{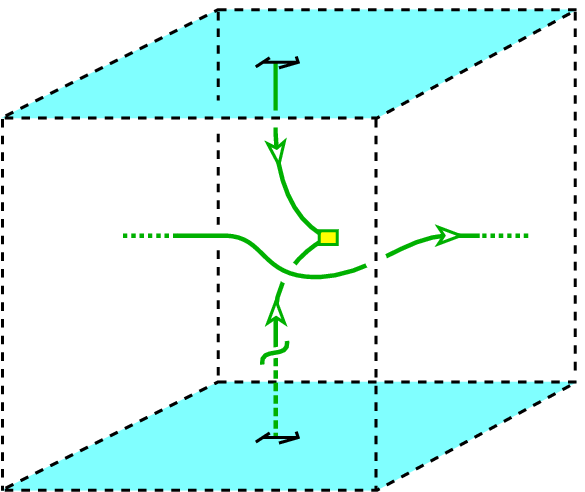}}}
   \put(0,0){
     \setlength{\unitlength}{.75pt}\put(-5,-3){
     \put (132,82) {\scriptsize$ 2 $}
     \put (75,99) {\scriptsize$ 3 $}
     \put (74,50) {\scriptsize$ 3 $}
     }\setlength{\unitlength}{1pt}}
  \end{picture}}
  \Bigg)
\ee
and the right hand side translates to
\be
  R = C_{[\Phi]} \, e^{\pi i h_3} \big(\eta^{(2)}_\phi\big)^2 \,
  \Bigg(   
  \raisebox{-35pt}{
  \begin{picture}(130,105)
   \put(0,0){\scalebox{.75}{\includegraphics{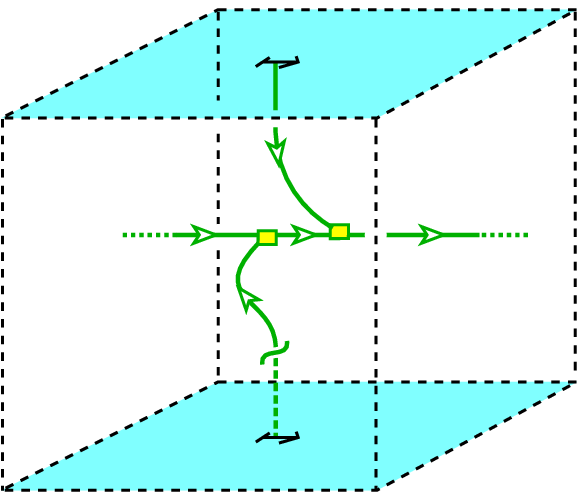}}}
   \put(0,0){
     \setlength{\unitlength}{.75pt}\put(-5,-3){
     \put (132,82) {\scriptsize$ 2 $}
     \put (90,65) {\scriptsize$ 2 $}
     \put (55,82) {\scriptsize$ 2 $}
     \put (89,96) {\scriptsize$ 3 $}
     \put (74,46) {\scriptsize$ 3 $}
     }\setlength{\unitlength}{1pt}}
  \end{picture}}
  ~-~
  \raisebox{-35pt}{
  \begin{picture}(130,105)
   \put(0,0){\scalebox{.75}{\includegraphics{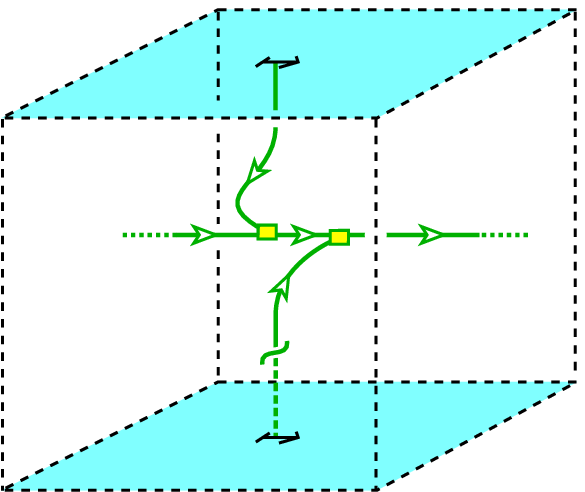}}}
   \put(0,0){
     \setlength{\unitlength}{.75pt}\put(-5,-3){
     \put (132,82) {\scriptsize$ 2 $}
     \put (90,82) {\scriptsize$ 2 $}
     \put (55,82) {\scriptsize$ 2 $}
     \put (87,96) {\scriptsize$ 3 $}
     \put (76,55) {\scriptsize$ 3 $}
     }\setlength{\unitlength}{1pt}}
  \end{picture}}
  \Bigg)
\ee
The constant $C_{[\Phi]}$ is determined by
$L = R$, i.e.\ by
\be
  \raisebox{-42pt}{
  \begin{picture}(70,85)
   \put(0,8){\scalebox{.75}{\includegraphics{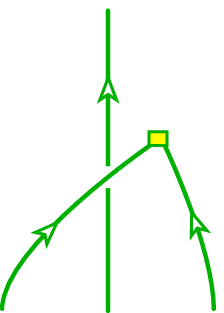}}}
   \put(0,8){
     \setlength{\unitlength}{.90pt}\put(-16,-16){
     \put( 22,  20) {\scriptsize$ 3 $}
     \put( 45,  30) {\scriptsize$ 2$}
     \put( 60,  20) {\scriptsize$ 3 $}
     }\setlength{\unitlength}{1pt}}
  \end{picture}}
 \hspace{-1em} - ~
  \raisebox{-42pt}{
  \begin{picture}(70,85)
   \put(0,8){\scalebox{.75}{\includegraphics{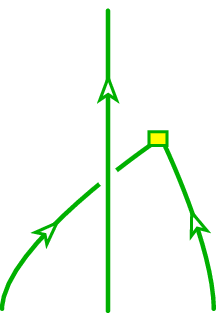}}}
   \put(0,8){
     \setlength{\unitlength}{.90pt}\put(-16,-16){
     \put( 22,  20) {\scriptsize$ 3 $}
     \put( 45,  30) {\scriptsize$ 2$}
     \put( 60,  20) {\scriptsize$ 3 $}
     }\setlength{\unitlength}{1pt}}
  \end{picture}}
  \hspace{-1em}
  \overset{?}{=} ~ \frac{ \big(\eta^{(2)}_\phi\big)^2}{\dim(3) \, \eta_\Phi} C_{[\Phi]}~
\Bigg(~
  \raisebox{-42pt}{
  \begin{picture}(70,85)
   \put(0,8){\scalebox{.75}{\includegraphics{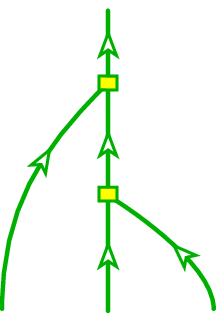}}}
   \put(0,8){
     \setlength{\unitlength}{.90pt}\put(-16,-16){
     \put( 18,  20) {\scriptsize$ 3 $}
     \put( 46,  20) {\scriptsize$ 2$}
     \put( 69,  20) {\scriptsize$ 3 $}
     \put( 46,  55) {\scriptsize$ 2$}
     \put( 46,  80) {\scriptsize$ 2$}
     }\setlength{\unitlength}{1pt}}
  \end{picture}}
 \hspace{-1em} - ~
  \raisebox{-42pt}{
  \begin{picture}(70,85)
   \put(0,8){\scalebox{.75}{\includegraphics{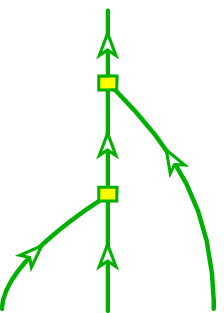}}}
   \put(0,8){
     \setlength{\unitlength}{.90pt}\put(-16,-16){
     \put( 24,  20) {\scriptsize$ 3 $}
     \put( 46,  20) {\scriptsize$ 2$}
     \put( 69,  20) {\scriptsize$ 3 $}
     \put( 33,  55) {\scriptsize$ 2$}
     \put( 46,  80) {\scriptsize$ 2$}
     }\setlength{\unitlength}{1pt}}
  \end{picture}}
\hspace{-1em} \Bigg)~  .
\labl{eq:comm-Phi-aux1}
Both sides represent vectors in a two dimensional space of intertwiners, so it is not clear that the equation can be solved. To see that it can, and that the value of $C_{[\Phi]}$ is as in \eqref{eq:C-Phi-comm-const}, compose \eqref{eq:comm-Phi-aux1} from below with
\be
  \raisebox{-42pt}{
  \begin{picture}(70,85)
   \put(0,8){\scalebox{.75}{\includegraphics{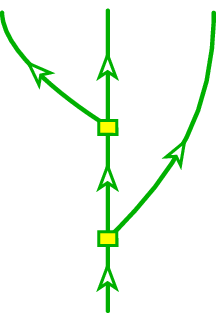}}}
   \put(0,8){
     \setlength{\unitlength}{.90pt}\put(-16,-16){
     \put( 22,  80) {\scriptsize$ 3 $}
     \put( 45,  80) {\scriptsize$ 2$}
     \put( 69,  80) {\scriptsize$ 3 $}
     \put( 33,  45) {\scriptsize$ k$}
     \put( 46,  20) {\scriptsize$ 2$}
     }\setlength{\unitlength}{1pt}}
  \end{picture}}
  ,
\ee
where $k \in \{ 2\equiv (1,2), 4 \equiv (1,4) \}$. Using \eqref{eq:local-ribbon-moves},  this results in
\be
  \Rs^{(23)k} \, \Fs^{(233)2}_{1k} - 
  \frac{1}{\Rs^{(23)k}}  \Fs^{(233)2}_{1k} 
  = 
  \frac{ \big(\eta^{(2)}_\phi\big)^2}{\dim(3) \, \eta_\Phi} C_{[\Phi]} \big(
  \Fs^{(323)2}_{2k} - \delta_{k,2} \big) \ .
\labl{eq:comm-Phi-aux2}
To continue, we need the explicit values from \eqref{eq:mm-F-2},
\be\bearll
 \Fs^{(233)2}_{12} = \frac{-1}{1{-}3t} \, \frac{\Ga{2{-}2t}\,\Ga{3t}}{\Ga{1{-}t}\,\Ga{2t}} ~~,\quad
 \etb
 \Fs^{(233)2}_{14} = \frac{\Ga{2{-}2t}\,\Ga{1{-}3t}}{\Ga{1{-}t}\,\Ga{2{-}4t}} \ ,
\\[1.5em]\displaystyle
 \Fs^{(323)2}_{22} = - \frac{\sin(\pi t)}{\sin(3 \pi t)} ~~,\quad
\etb
 \Fs^{(323)2}_{24} = \frac{\Ga{1{-}3t}\,\Ga{2{-}3t}}{\Ga{2{-}4t}\,\Ga{1{-}2t}} \ ,
\eear\ee
as well as \eqref{eq:mm-dim} and \eqref{eq:mm-Rs-val}.
After a short calculation one finds that for both values of $k$, \eqref{eq:C-Phi-comm-const} solves \eqref{eq:comm-Phi-aux2}.

\appendix

\sect{Appendix: Chiral data for minimal models}\label{app:mm-data}

In this appendix, various minimal model data used in the main text is collected.
Let $p,p'\ge 2$ be two coprime integers and set $t=p/p'$. 
For $a=(r,s)$ an entry in the Kac-table set
\be
  d_a = r-st \quad , \quad
  h_a = \tfrac1{4t}(d_a^{\,2} - d_1^{\,2}) \ .
\ee
Here and below we use the following abbreviations for entries in the Kac-table
\be
  1 = (1,1) ~~,~~~
  2 = (1,2)~~,~~~
  3 = (1,3) \ .
\ee
The set $\Ic$ denotes the set of Kac-labels modulo the identification $(r,s) \sim (p-r,p'-s)$.
We also use the notation $(r,s)$ for elements of $\Ic$. 

The modular matrix $S_{ab}$, which describes the behaviour of characters under the modular $S$-transformation, reads (see e.g.\ \cite[\S\,10.6]{DiFr:1997})
\be
  S_{(r,s)(x,y)} = \sqrt{8/(p p')} \, (-1)^{1+ry+sx} \, \sin( \pi rx/ t ) \sin( \pi s y  t) \ .
\ee
The quantum dimension $\dim(a)$ of the representation $R_a$ is, for $a = (r,s)$,
\be
  \dim(a) = \frac{S_{a1}}{S_{11}} =  
  (-1)^{r-1} \, \frac{\sin( \pi r / t )}{\sin( \pi / t )} \cdot
  (-1)^{s-1} \, \frac{\sin( \pi s t)}{\sin( \pi t)} \ .
\labl{eq:mm-dim}
Denote by $N_{ab}^{~c}$ the fusion
rule coefficients, i.e.\ the dimension of $\Hom(R_a \otimes R_b,R_c)$.
The braiding matrix is given by, for $a,b,c \in \Ic$ with $N_{ab}^{~c}=1$
(see e.g.\ \cite{Moore:1989vd})
\be
  \Rs^{a(bc)} = \exp(\pi i( h_b+h_c-h_a) ) \ .
\labl{eq:mm-Rs-val}
The $\Fs$-matrix entries can be determined via a closed form expression \cite{Dotsenko:1984ad,Furlan:1989ra} or recursively \cite{Runkel:1998pm}. 
The inverse $\Gs$ of the $\Fs$-matrix is given by
(combine \cite[eqn.\,(2.61)]{tft1} with \eqref{eq:mm-Rs-val})
\be
  \Gs^{(abc)d}_{pq} = \Fs^{(cba)d}_{pq}  ~.
\labl{eq:mm-GF}
The $\Fs$-matrix has the symmetries
\be
  \Fs^{(abc)d}_{pq}
  = \Fs^{(bad)c}_{pq}
  = \Fs^{(cda)b}_{pq}
  = \Fs^{(dcb)a}_{pq} \ ,
\ee
and obeys
\be
  \Fs^{(aaa)a}_{11} = \frac{1}{\dim(a)} \quad \text{and} \quad
  \Fs^{(abb)a}_{1p} \, \Fs^{(aab)b}_{p1} = \frac{\dim(p)}{\dim(a)\dim(b)} \ ,
\labl{eq:F1p-vs-Fp1}
see e.g.\ \cite{Moore:1989vd}.
Some special entries used in the main text are the following.
For $\eps,\nu = \pm 1$ and $a,b,c \in \Ic$, 
\begin{align}
  \Fs^{(a2b)c}_{b+\eps,a+\nu} 
  &= \frac{ \Ga{\nu d_a} \, \Ga{1{-}\eps d_b} }{
    \Ga{\tfrac12(1{+}d_c{+}\nu d_a{-}\eps d_b)}
    \Ga{\tfrac12(1{-}d_c{+}\nu d_a{-}\eps d_b)} } \ ,
\label{eq:mm-F-2}
\\[.5em]
  \Fs^{(33a)a}_{a1} &= 
  \frac1{1{-}3t}\,
  \frac{
  \Ga{1{-}t{+}d_a}\, \Ga{1{-}t{-}d_a} \,\Ga{1{-}t} \,\Ga{2t}\, \Ga{3t}
  }{
  \Ga{t{-}d_a}\, \Ga{t{+}d_a}\, \Ga{2{-}2t}\, \Ga{t}^2
  } \ ,
\label{eq:F(33a)a,a1)}
\\[.5em]
  \Fs^{(33a)a}_{a3} &=
  - \frac{2 \sin(\pi t) \cos( \pi d_a ) }{ \sin(4 \pi t) }
  \frac{
  \Ga{1{-}t{+}d_a}\, \Ga{1{-}t{-}d_a}}{\Ga{2{-}4t}\, \Ga{2t}} \ .
\label{eq:F(33a)a,a3}
\end{align}
For $p'{-}1 \equiv (1,p'{-}1)$, $s \equiv (1,s)$ and $p'{-}s \equiv (1,p'{-}s)$,
\be
  \Fs^{(p'-1,s,3)p'-s}_{s,p'-s}
  = (-1)^p \,
  \frac{ \Ga{2{-}(s{+}1)t} \, \Ga{(s{-}1)t} }{
  \Ga{2{-}(p'{-}s{+}1)t} \, \Ga{(p'{-}s{-}1)t} } ~.
\labl{eq:F(ps3)}
Here it is understood that the indices are chosen such that the $\Fs$-matrix entry is
well-defined (i.e.\ the corresponding spaces of conformal blocks are non-vanishing, 
see e.g.\ \cite[App.\ A.1]{Runkel:2007wd} for more details using the present notation).

\end{document}